\documentclass[12pt,preprint]{aastex}

\newcommand{\ASCA}{{\it ASCA}\xspace}

\newcommand{\HIPPARCOS}{{\it HIPPARCOS}\xspace}
\newcommand{\EINSTEIN}{{\it Einstein}\xspace}

\newcommand{\ROSAT}{{\it ROSAT}\xspace}
\newcommand{\XMM}{{\it XMM-Newton}\xspace}

\newcommand{\CHANDRA}{{\it Chandra}\xspace}

\newcommand{\XRT}{{XRT}\xspace}

\newcommand{\UNITFLUX}{{\rm ergs~cm$^{-2}$~s$^{-1}$}\xspace}
\newcommand{\UNITCPS}{{\rm cnts~s$^{-1}$}\xspace}

\newcommand{\UNITCPPIX}{{\rm cnts~pixel$^{-1}$}\xspace}

\newcommand{\UNITLUMI}{{\rm ergs~s$^{-1}$}\xspace}

\newcommand{\UNITAREA}{{\rm cm$^{-2}$}\xspace}
\newcommand{\UNITNH}{{\rm cm$^{-2}$}\xspace}
\newcommand{\UNITVEL}{{\rm km~s$^{-1}$}\xspace}
\newcommand{\UNITSOLARLUMI}{{\it L$_{\odot}$}\xspace}
\newcommand{\UNITSOLARMASS}{{\it M$_{\odot}$}\xspace}

\newcommand{\UNITSOLARABUND}{{\it Z$_{\odot}$}\xspace}

\newcommand{\DEGREEKELV}{K\xspace}
\newcommand{\DEGREE}{{$^{\circ}$}\xspace}
\newcommand{\ARCMIN}{{$'$}\xspace}
\newcommand{\FARCM}{{.$'$}}
\newcommand{\ARCSEC}{{$''$}\xspace}
\newcommand{\FARCS}{.{$''$}}
\newcommand{\NH}{{\it N$_{\rm H}$}\xspace}
\newcommand{\AV}{{\it A$_{\rm V}$}\xspace}
\newcommand{\RV}{{\it R$_{\rm V}$}\xspace}

\newcommand{\LX}{{\it L$_{\rm X}$}\xspace}
\newcommand{\Lbol}{{\it L$_{\rm bol}$}\xspace}

\newcommand{\Teff}{{\it T$_{\rm eff}$}\xspace}

\newcommand{\KT}{{\it kT}\xspace}

\newcommand{\EM}{{\it E.M.}\xspace}

\newcommand{\PSF}{{\it psf}\xspace}
\newcommand{\FOV}{{\it fov}\xspace}

\newcommand{\HAEBE}{HAeBe\xspace}
\newcommand{\HAEBEs}{HAeBes\xspace}

\newcommand{\Mdot}{{$\dot M$}\xspace}

\usepackage{xspace}
\usepackage[]{natbib}
\shorttitle{X-ray Study of the Herbig Ae/Be stars}
\shortauthors{Hamaguchi et al.}

\begin{document}

\title{X-ray Study of the Intermediate-Mass Young Stars Herbig Ae/Be Stars}

\author{Kenji Hamaguchi\altaffilmark{1}}
\affil{Laboratory for High Energy Astrophysics, NASA/Goddard Space Flight Center,
Greenbelt, MD 20771, USA}
\email{kenji@milkyway.gsfc.nasa.gov}

\author{Shigeo Yamauchi}
\affil{Faculty of Humanities and Social Sciences, Iwate University, 3-18-34 Ueda,
Morioka, Iwate 020-8550, Japan}
\email{yamauchi@iwate-u.ac.jp}
\and

\author{Katsuji Koyama}
\affil{Department of Physics, Faculty of Science, Kyoto University\\
Kitashirakawa-oiwakecho, Sakyo, Kyoto 606-8502, Japan}
\email{koyama@cr.scphys.kyoto-u.ac.jp}

\altaffiltext{1}{National Research Council, 500 Fifth Street, NW, Washington, DC 20001, USA}

\begin{abstract}
We present the \ASCA results of intermediate-mass pre-main-sequence
stars (PMSs), or  Herbig Ae/Be stars (\HAEBEs).
Among the 35 \ASCA pointed-sources,
we detect 11 plausible X-ray counterparts.
X-ray luminosities of the detected sources in the 0.5--10 keV band are 
in the range of log \LX $\sim$30$-$32 \UNITLUMI, which is systematically
higher than those of low-mass PMSs.
This fact suggests that the contribution of a possible low-mass companion
is not large.
Most of the bright sources show significant time variation, particularly, two \HAEBEs~- MWC~297 and TY~CrA - exhibit flare-like
events with long decay timescales ($e$-folding time $\sim$ 10$-$60 ksec).
These flare shapes are similar to those of low-mass PMSs.
The X-ray spectra are successfully reproduced by an absorbed 
one or two-temperature thin-thermal plasma model.
The temperatures are in the range of \KT $\sim$ 1$-$5 keV, which are
significantly higher than those of main-sequence OB stars (\KT $<$ 1 keV).
These  X-ray properties are not explained by wind driven shocks, 
but are  more likely due to magnetic activity.
On the other hand, 
the plasma temperature rises as absorption
column density increases, or as \HAEBEs ascend to earlier phases.
The X-ray luminosity reduces after stellar age of a few$\times$10$^{6}$ years.
X-ray activity may be  related to stellar evolution.
The age of the activity decay is apparently near the termination of jet or outflow activity.
We thus hypothesize that magnetic activity originates from the interaction of the 
large scale magnetic fields coupled to the circumstellar disk.
We also discuss differences in X-ray properties between \HAEBEs and main-sequence OB stars.
\end{abstract}


\section{Introduction}
Main sequence stars (MSs)  in the intermediate-mass range
(2$-$10 \UNITSOLARMASS) exhibit no significant X-ray activity \citep[e.g.][]{Rosner1985,Berghoefer1997}, due  to the absence of X-ray production
mechanisms: neither strong UV field to accelerate high-speed stellar 
winds working in high-mass stars, nor surface convection responsible for magnetic activity working in low-mass stars. Intermediate and high-mass stars in the pre-main-sequence (PMS) stage, called Herbig Ae/Be stars
\cite[\HAEBEs,][]{Herbig1960,Waters1998}
have also neither strong UV fields nor 
surface convection zones, hence no X-ray emission is predicted for them.
Nevertheless the \ROSAT and \EINSTEIN surveys detected the X-ray emission from 
significant numbers of \HAEBEs \citep{Zinnecker1994,Damiani1994}.
\citet{Damiani1994}, using \EINSTEIN, detected X-ray emission from 11 \HAEBEs out of 31 samples. They found that Be stars are more luminous than Ae stars, and
suggested that the X-ray luminosity correlates  to the terminal 
wind velocity, but not  to the stellar rotational velocity $v_{rot}$ sin $i$.
\citet{Zinnecker1994} detected 11 \HAEBEs out of 21 \ROSAT samples.
In contrast to the \EINSTEIN result, they  found no 
significant  difference of X-ray luminosity between Be and Ae stars.
They found no correlation of X-ray luminosity to $v_{rot}$ sin $i$,
but found a correlation to mass loss rate.
Thus both the \ROSAT and \EINSTEIN results suggested that X-ray emission of \HAEBEs relates to the  stellar winds.

The  \ASCA spectra from  some bright \HAEBEs (HD 104237, IRAS 12496$-$7650 and MWC 297) exhibited high temperature plasma ({\it kT} $\sim$3 keV), while  MWC 297 exhibited a large X-ray flare \citep {Skinner1996,Yamauchi1998,Hamaguchi2000}. 
Such hot plasma and/or rapid X-ray variability can not be produced by 
stellar wind.
Such properties are usually seen in low-mass stellar X-rays originating from magnetic activity.
In fact, magnetic activity of
\HAEBEs  may be found in the powerful outflows and jets \citep{Mundt1994}.

Wide band imaging spectroscopy between 0.5$-$10 keV first enabled with \ASCA is a powerful tool to 
measure plasma temperatures and time variations.
We thus perform a survey of  \HAEBEs using the \ASCA archive.
From the systematic X-ray spectrum and timing analysis,
we study the origin of the X-ray emission from \HAEBEs 
with reference to the \ROSAT and \EINSTEIN results.
The combined analysis leads us to an overall picture across the whole 
pre-main-sequence (PMS) phase of low and intermediate-mass stars.

This paper comprises as follows. The method of target selection is described in Section 2.
Data analysis and brief results are summarized in Section 3.
Comments on individual sources are in Section 4.
The discussion of the X-ray emission mechanism and related phenomena
to MS stars are described in Section 5 and 6.
Section 7 summarizes results.

\section{Target Selection \& Data Reduction}
We obtained X-ray data from 6 HAeBes, V892~Tau, IRAS 12496$-$7650, MWC~297,
HD 176386, TY CrA, and MWC~1080 with our \ASCA guest observing programs.
We further picked up HAeBe samples from the \ASCA
archive using the comprehensive catalog compiled by \citet{The1994} (above
spectral type F in Table i, ii and v) and the recent lists by 
\citet{Ancker1997,Ancker1998}.
We also analyzed the proto-HAeBe EC95 \citep{Preibisch1999}, that
is not included in the statistical analysis of \HAEBE X-ray properties but used
for evolutional study of the X-ray emission.
The list of the observed \HAEBEs is given in Table \ref{tbl:starpar}.
Optical positions in Column 2, 3 and distance in Column 4 are from
\HIPPARCOS
if available \citep{Ancker1997,Ancker1998,Bertout1999}.
The \ASCA observation log is shown in Table \ref{tbl:ascaobslog}.

The fourth Japanese X-ray satellite \ASCA, launched in 1993, has four multi-nested 
Wolter type I X-ray telescopes (\XRT),
on whose focal planes two X-ray CCD cameras (Solid-state Imaging Spectrometer: SIS) and two gas scintillation proportional counters 
(Gas Imaging Spectrometer: GIS) are installed \citep{Tanaka1994}.
The XRTs have a cusped shape point spread function (\PSF) with the half power diameter of $\sim$3\ARCMIN,
which is slightly degraded for the GISs by their detector response.
The effective area of the four sets of mirrors are around 1400 \UNITAREA at 2 keV and 900 \UNITAREA at 5 keV.
Bandpasses of the SIS and the GIS combined with the XRT are around 0.4$-$10 keV 
and 0.7$-$10 keV, respectively.
The SISs have moderate spectral resolution of $\sim$130 eV at 5.9 keV and 
field of view (\FOV) of 20\ARCMIN $\times$20\ARCMIN with four CCD chips fabricated 
on each SIS \citep{Burke1994,Yamashita1999}.
A CCD chip needs 4 seconds to be read out, and runs of 1, 2 and 4 chips
(corresponding to 1, 2 and 4 CCD modes) need 4, 8 and 16 seconds in an exposure, respectively.
%% Run of all CCD chips needs longer frame exposure time ($\sim$16 sec) than run of 
%% one or two CCD chips (4, 8 sec).
After $\sim$1995, spectral resolution of data taken with 2 and 4 CCD modes significantly degraded, 
which is caused by development of pixel-to-pixel variation of dark current 
(residual dark distribution: RDD).
The clocking and telemetry modes of the SIS detectors are summarized in Column 5.
The GISs have energy resolution of $\sim$460 eV at 5.9 keV and wide \FOV (50\ARCMIN diameter)
\citep{Ohashi1996,Makishima1996}.
All observations  except for PSR J0631 were made with the standard GIS (PH) mode,
while a non-standard  bit assignment (9-8-8-0-0-0-7) was used for PSR J0631.

We use the \ASCA "revision2" archival data.
The data basically do not need further reduction, but
the degradation of the SIS data by RDD in the 2 and 4 CCD modes are restored
using {\it correctrdd} \citep{Dotani1997}.
We remove hot and flickering pixels in the SIS using {\it sisclean}
and select the \ASCA grade 0234.  The GIS events are selected using {\it gisclean} \citep{Ohashi1996}.
Both the SIS and the GIS data are screened using the standard  criteria,
which excludes data in the South Atlantic Anomaly, Earth occultation and high-background regions with low-geomagnetic rigidities.
The SIS data with viewing angle less than 20\DEGREE above  the day-earth horizon
are also excluded. 
Analyses are made using the software package FTOOLS 4.2,
XIMAGE ver.~2.53, XSPEC ver.~9.0 and XRONOS ver.~4.02.
%%For the imaging of the sources near the GIS outer rim, we applied the Display45 (ver.~1.90) package \citep{Ishisaki1997}.

\section{Analysis \& Results}

\subsection{Source Detection}
\label{subsec:stellaridentifications}

The analysis is basically made in the 0.5--10 keV band for the SIS and 
the 0.8--10 keV band for the GIS.
The low energy threshold for the SIS is set at slightly higher level than those 
of  the onboard discrimination because
the low energy efficiency at these levels are  degraded by RDD and the 
event splitting effects can not be fully corrected by {\it correctrdd}.

We made X-ray images for both the SIS and the GIS in J2000 coordinate,
where the absolute coordinate is corrected by the method in \citet{Gotthelf2000}.
The positional accuracy for  the SIS bright sources should be within $\sim$12\ARCSEC.
We searched for X-ray sources above 5$\sigma$ detection level around the position of relevant \HAEBE.  Still possible mis-identification to nearby sources can not be excluded in crowded regions. 
We checked the \ROSAT results in \citet{Zinnecker1994} and \citet{Preibisch1998} and/or retrieved the \ROSAT images (Table \ref{tbl:rosatobslog}) to confirm the source identifications.
To extract the X-ray  events for spectral and timing analysis, 
we normally took a circle of 2\FARCM5 or
3\ARCMIN radius around the relevant source. 
The background region is selected from the nearby source-free region or at  the symmetrical region with respect to a contaminating source, if any.
The selected background regions  are summarized in Table \ref{tbl:reg}.

For non-detected \HAEBEs, we determine the 3$\sigma$ upper-limit (99\% confidence levels) from a 40\ARCSEC 
square region for the SIS and a 80\ARCSEC square region for the GIS.  
The background data are taken from a nearby source-free region with the {\it sosta} package in XIMAGE.
The upper-limits are converted to fluxies using the PIMMS package
assuming an absorbed thin-thermal model with \KT = 1 keV, abundance = 0.3 \UNITSOLARABUND  and \NH converted from \AV \citep[\NH $\sim$2.2 $\times$10$^{21}$\AV  \UNITNH,][]{Ryter1996}.
The upper-limit flux for retrieving \ROSAT data is
estimated using  the same method as the \ASCA data, except that
the background is taken  from a surrounding 12\ARCMIN$\times$12\ARCMIN region.
\ROSAT has narrower bandpass (0.1--2.4 keV) than \ASCA, but it has sharper on-axis \PSF, and
therefore is better for source identification especially in a crowded region.

Table \ref{tbl:imgdet} and \ref{tbl:imgnondet} list the detected and non
detected sources, respectively.
With the \ASCA satellite, 11 sources are found above the 5 $\sigma$ threshold.
(V921 Sco is excluded due to difficulty in identification.)
The detection rate ($\sim$ 31\%) is smaller than that of 
the \ROSAT survey ($\sim$ 52\%, \citealt{Zinnecker1994}), due mainly to the higher detection threshold to eliminate contamination from nearby bright sources (e.g. AB Aur, HD 97048 and Z CMa).
A few embedded sources (e.g. MWC 297 and IRAS 12496-7650) are however 
detected for the first time in X-rays by \ASCA, thanks to its hard X-ray imaging 
capability.

\subsection{Timing Analysis}
\label{subsec:summarylightcurves}

We made light curves in the full energy range by subtracting background. 
The  time bin is typically 2048 sec but it is longer for weak sources.
The error of each bin is estimated by the Gaussian approximation.
Typical net counts per bin of combined SIS and GIS light curves 
are around 80 counts/bin.
We fit the light curves with a constant-flux model using the chi-square method, 
and find time variable sources with 96\% confidence.

Table \ref{tbl:curvefit} shows the result of the timing analyses.
Four (or five if we include merged light curves of TY CrA and HD 176386 (TYHD)) among twelve detected sources  (36--45\%) were variable.
If we exclude sources with total net counts less than 1000 
photons, half (or 3/4 including TYHD) are variable.

\subsection{Spectra}
\label{subsec:summaryspectra}
\label{subsec:SpectralAnalysis}

We fit spectra with  an absorbed thin-thermal plasma model (the MeKaL plasma code, \citealt{Mewe1995}).
The response matrixes for the SIS were made with {\it sisrmg} ver.~1.1, while a standard one
was used for the GIS.
The ancillary response function for both were made with {\it ascaarf} ver.~2.81.
Chemical abundances are basically fixed at 0.3\UNITSOLARABUND, following the typical value
for the  stellar X-rays (e.g. OB stars, \citealt{Kitamoto1996,Kitamoto2000}; low-mass MSs, \citealt{Tagliaferri1997}; low-mass PMSs, 
\citealt{Yamauchi1996,Kamata1997,Tsuboi1998}).

Spectra for most sources except HD 104237 and HD 200775 were fit  with either  an absorbed one-temperature (1T)
or two-temperature (2T) thin-thermal (MeKaL) model (with a Gaussian component for VY Mon) with 90\% confidence (Table \ref{tbl:specfit}).
Unacceptable fits of HD 104237 and HD 200775 seem to be caused by small structures in the spectra,
which might be unknown emission lines or absorptions.
The best-fit temperature ranges between \KT $\sim$1$-$5 keV, which is significantly
higher than the X-ray temperature of MS OB stars \citep[\KT $<$1 keV, e.g. ][]{Corcoran1994}.

\section{Comments on Individual Sources}
This section provides observational details of each detected source: i.e. stellar characteristics and \ASCA
results of imaging, timing and spectral analysis.
%%Sources which are sometimes mentioned as \HAEBEs but with the spectral type less than G are
%%summarized in the latter of this section.

\subsection{V892 Tau}
\label{sec:v892tau}
V892 Tau (Elias 3-1) is located in the L1495E dark cloud.
The stellar parameters are open to dispute;
\citet{Strom1994} quoted it as a B9 star of \Lbol$\sim$64.3 \UNITSOLARLUMI, 
while \citet{Zinnecker1994} quoted it as an A6e star of 
\Lbol$\sim$38 \UNITSOLARLUMI.
It has a circumstellar disk of $\sim$0.1 \UNITSOLARMASS
\citep{DiFrancesco1997}.
There is a companion star (Elias I NE) at $\sim$4\ARCSEC northeast
from V892 Tau, which would be a weak line T-Tauri star (WTTS) with \Lbol$\sim$0.3 \UNITSOLARLUMI
\citep{Skinner1993,Leinert1997,Pirzkal1997}.
\citet{Strom1994} first reported the X-ray emission from V892 Tau
(log \LX $\sim$30 \UNITLUMI).
\citet{Zinnecker1994} measured a plasma temperature of
$\sim$2.3 keV and hydrogen column density of
$\sim$4.8 $\times$10$^{21}$ \UNITNH.

\begin{figure}
\plotone{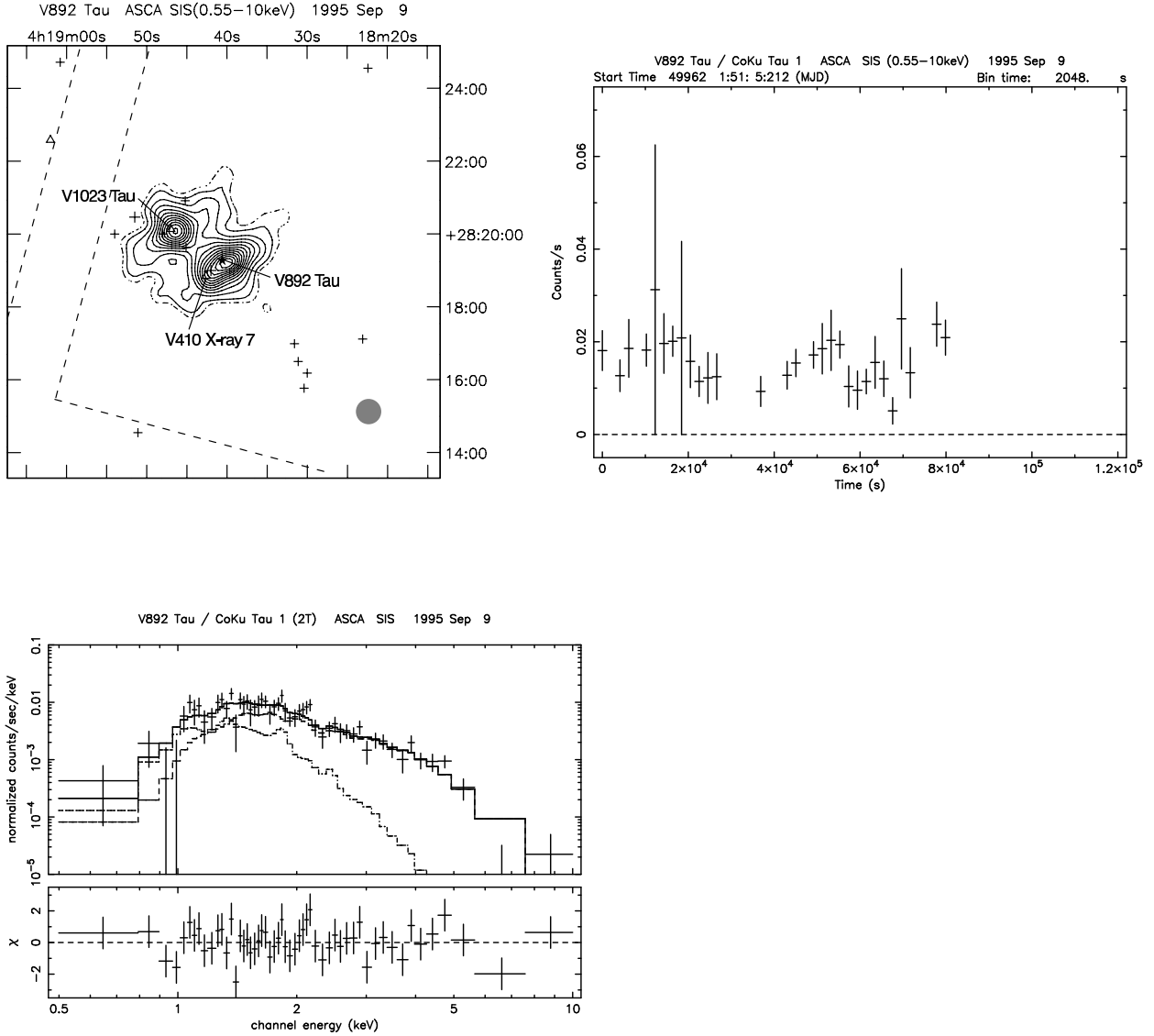}
\caption{a: \ASCA image of the V892~Tau field (12\ARCMIN $\times$12\ARCMIN) ({\it top left panel}).
The contour image smoothed with a Gaussian function of $\sigma\sim$2 pixel
(SIS: 6\FARCS4 pixel$^{-1}$, GIS: 14\FARCS7 pixel$^{-1}$)
is drawn by 0.4 \UNITCPPIX for the SIS and 0.5 \UNITCPPIX for the GIS.
A dot-dash line represents 5$\sigma$ detection levels of 
the Poisson-distributed photon error of the background.
Dash-lines are the detector rims.
The filled circle on the bottom right shows the typical 
error circle of the SIS (20\ARCSEC) and the GIS (40\ARCSEC). The coordinate system is {\it J2000}.
{\it Legend $-$} star: \HAEBE, diamond: infrared source, triangle: young star and PMS object, square: radio source, cross: normal star, asterisk: galaxy. The source category is referred to the SIMBAD database.
b: SIS light curves of V892 Tau ({\it top right panel}). The background is subtracted.
The vertical axis is averaged detector count rates.
Dotted lines are the zero level. 1 bin is 2,048 sec.
c: SIS spectrum of V892~Tau including the component of V410~X-ray~7 (2T model in Table \ref{tbl:specfit}, 
{\it bottom left panel}).
The solid and dotted lines show the best-fit model of the total and
each component, respectively.
The residual of the best-fit model is displayed in the lower section.
\label{fig:v892tau}}
\end{figure}

V892 Tau is detected as the strongest source in the 
12\ARCMIN $\times$12\ARCMIN square field (Figure \ref{fig:v892tau}a).
The SIS image shows two other X-ray sources: V1023 Tau 
at $\sim$2\ARCMIN northeast from V892 Tau and V410~X-ray~7
at 30\ARCSEC southeast, which is only obviously seen in the soft band image
(0.55$-$2 keV).
V410~X-ray~7 is a reddened M0.5 PMS star \citep{Briceno1998},
from which \ROSAT detected an X-ray flare \citep{Stelzer2000}.
Because the GIS does not resolve V1023 Tau from V892 Tau,
we only use the SIS for latter analysis.
V892 Tau and V410~X-ray~7 strongly merge together
so that V410~X-ray~7 is included in the source region of V892 Tau.
The light curve in the total band (Figure \ref{fig:v892tau}b)
seems to fluctuate by a factor of two, but it accepts a constant model
within 90\% confidence level, which suggests that neither V892 Tau nor V410~X-ray~7
showed variation.
The spectrum 
can be reproduced by an absorbed 1T thin-thermal model (1T model in Table \ref{tbl:specfit}),
but it includes a component from V410~X-ray~7.
As we see in the SIS hard band image, 
hard X-ray emission from V410~X-ray~7 is negligible.
Because the high energy part of the X-ray emission determines the plasma temperature,
we think that the temperature derived in the 1T model represents the temperature of V892 Tau.
On the other hand, 
in the soft band image (0.55$-$2 keV), V410~X-ray~7 is as 
bright as V892 Tau, which means that the soft X-ray flux of V892 Tau is half of that of the 1T model.
When we only change \NH parameter of the 1T model to fit the condition, 
\NH of V892 Tau is $\sim$1.8$\times$10$^{22}$ \UNITNH.
The residual component, which would be attributed to V410~X-ray~7, can be 
successfully fit by another absorbed 1T model (2T model in Table \ref{tbl:specfit}
and Figure \ref{fig:v892tau}c).

\subsection{V380 Ori}
\label{subsec:L1641obs}
V380 Ori is a B9--A0 star of 3.3$-$3.6 \UNITSOLARMASS with small 
amplitude variability
\citep{Hillenbrand1992,Boehm1995,Rossi1999,Herbst1999,Rossi1999}.
It has a companion star 0\FARCS15 from the primary \citep{Corporon1999}.
X-ray emission was detected with \EINSTEIN and \ROSAT 
(log \LX $\sim$ 31.0$-$31.3 \UNITLUMI, \citealt{Pravdo1981,Zinnecker1994}).
From the \EINSTEIN data, 
the plasma temperature and column density are 
$\sim$1.1 keV and 7.4$\times$10$^{21}$ \UNITNH, respectively
\citep{Damiani1994}.

\begin{figure}
\plotone{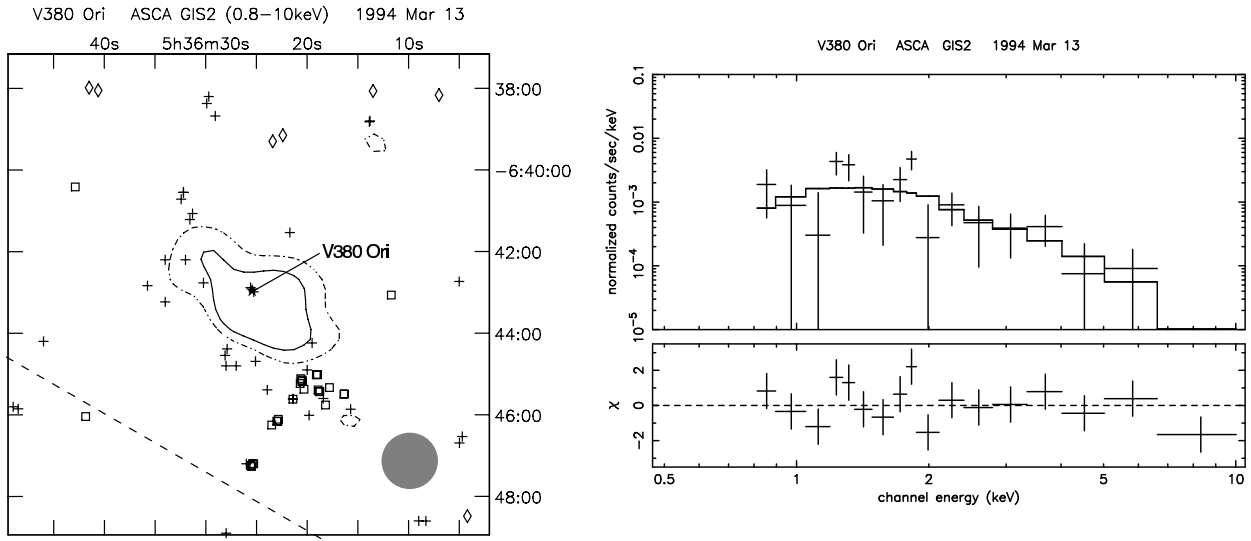}
\caption{a: GIS image of the V380~Ori field ({\it left panel}). See Figure \ref{fig:v892tau}a
for details.
b: GIS spectrum of V380 Ori ({\it right panel}). See Figure \ref{fig:v892tau}c for details.
\label{fig:v380ori}}
\end{figure}

V380 Ori is detected as a relatively faint source (Figure \ref{fig:v380ori}a).
It is near the edge of the GIS \FOV (4\ARCMIN inside from the edge) so that the \PSF is less sharp.
The light curve shows no apparent variability.
The spectral fitting (see Figure \ref{fig:v380ori}b) shows
relatively high plasma temperature ($\sim$3.2 keV)
though the uncertainty (0.7 $-$ 11.0 keV) is large.
The luminosity (log \LX $\sim$31.2 \UNITLUMI) is about the same as 
that in the \EINSTEIN and \ROSAT observations.

\subsection{VY Mon \& CoKu VY Mon/G2}
Both VY Mon and CoKu VY Mon/G2 are members of the Mon OB1 
association \citep{Cohen1979,Testi1998}.
VY Mon is a O9$-$B8 star in very young phase
($\sim$5$\times$10$^{4}$ years) with
Algol-type variability and large IR excess
\citep[see][]{Casey1990,The1994,Testi1999}.
CoKu VY Mon/G2 is a candidate Herbig Ae star 
\citep{The1994}.
\citet{Damiani1994} reported X-ray emission from
VY Mon and/or CoKu VY Mon/G2 with \EINSTEIN in the hard band
(S/N $\sim$2.5).

\begin{figure}
\plotone{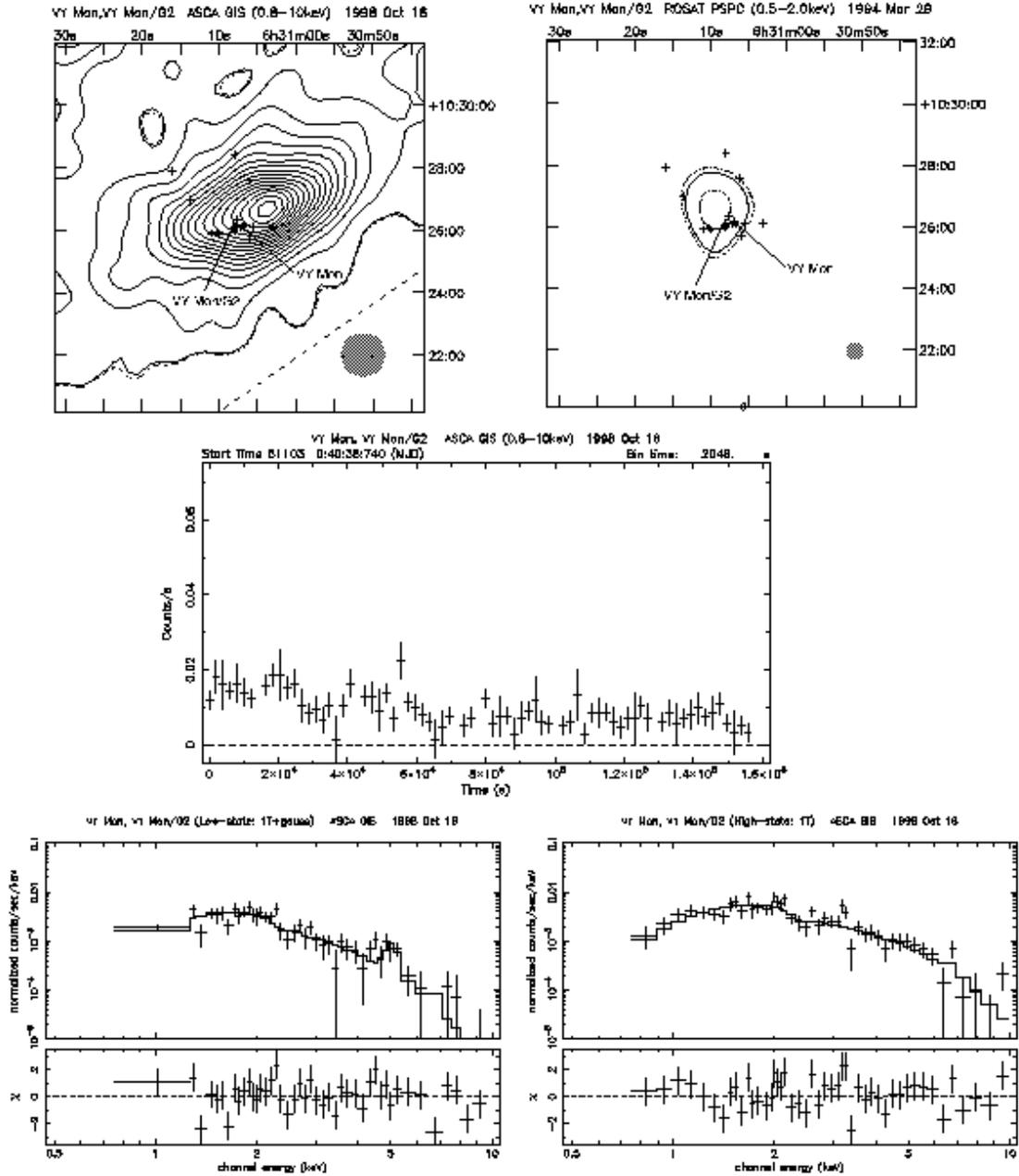}
\caption{a: \ASCA GIS image of the VY Mon field ({\it top left panel}) and 
\ROSAT PSPC image of the VY Mon field ({\it top right panel}).
The star marks are VY Mon ({\it right}) and CoKu VY Mon/G2 ({\it left}), respectively.
b: GIS light curves of VY Mon ({\it middle panel}).
c: GIS spectra of VY Mon ({\it bottom left panel}: LS, {\it bottom right panel}: HS).
\label{fig:vymon}}
\end{figure}

The GIS detects an X-ray source around VY Mon and CoKu VY Mon/G2
at a large off-axis angle ($\sim$20\ARCMIN), where the absolute position uncertainty is
poorly calibrated (a few arc-minutes) (Figure \ref{fig:vymon}a {\it left}).
We thus used a \ROSAT PSPC image for source identification (Figure \ref{fig:vymon}a {\it right}).
Though all four \ROSAT observations has relatively large off-axis angle of the source 
($\sim$30\ARCMIN), where 68\% of source photons falls within a $\sim$1\ARCMIN radius circle,
VY Mon and/or CoKu VY Mon/G2 can be the counterpart of the X-ray source.
The \ASCA light curve (Figure \ref{fig:vymon}b) gradually decreases with small 
fluctuations and becomes almost constant after $\sim$6$\times$10$^{4}$ sec.
We thus define the phase before 6$\times$10$^{4}$ sec as the high state (HS)
and the phase after as the low state (LS).
The low state spectrum has an enhancement at around 5 keV 
({\it left panel} of Figure \ref{fig:vymon}c).
An absorbed 1T model with a Gaussian component at $\sim$5.1 keV 
has an acceptable fit in 90\% confidence level, but there is no corresponding emission line
of abundant elements at the measured energy (e.g. Ca, Fe).
We also tried an additional thermal model, but it does not reproduce the strong dip feature
at $\sim$5.3 keV.
The origin of the line emission is unknown.
On the other hand, the high state spectrum ({\it right panel} of Figure \ref{fig:vymon}c)
accepts an absorbed 1T model with very high plasma temperature ($\sim$6 keV).
The X-ray luminosity (log \LX $\sim$$32.2$ \UNITLUMI)
is one of the largest among our samples.
VY Mon forms a small cluster. The X-ray emission could
be from the assembly of cluster members.
However, we should note that \NH does not change between the HS and LS, 
and corresponds to the \AV ($\sim$1.6$^{m}$) of CoKu VY Mon/G2.

\subsection{HD 104237}
HD 104237, located in the Chamaeleon III dark cloud,
is the brightest \HAEBE star in the sky \citep[V =6.6$^m$,][]{Hu1991}.
It has the spectral type A4e and would be 
slightly older than other star \HAEBE stars 
\citep[Age $\sim$10$^{6.3}$ years,][]{Knee1996,Ancker1998}.
The \ASCA observation of HD 104237 is described in \citet{Skinner1996} in detail.

\begin{figure}
\plotone{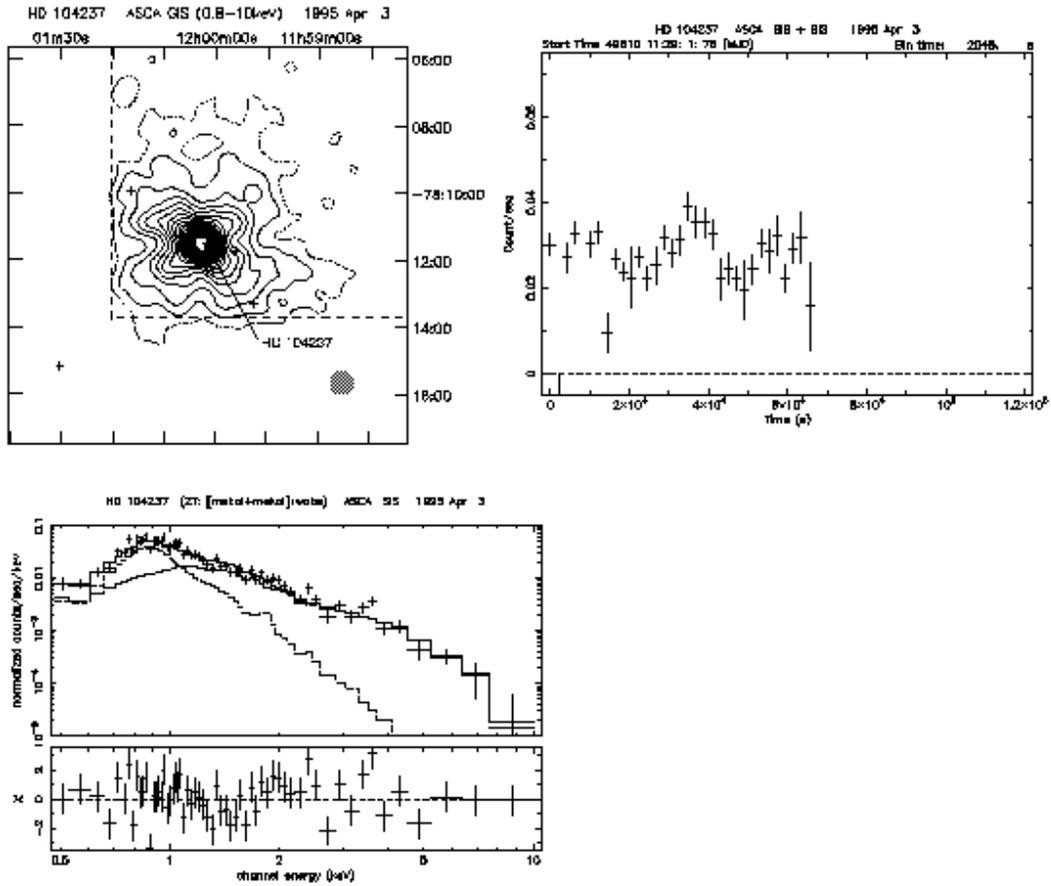}
\caption{a: SIS image of HD 104237 ({\it top left panel}).
b: Light curves of HD 104237 ({\it top right panel}).
c: SIS spectrum of HD 104237 fit by a commonly absorbed 2T model 
(2T model in Table \ref{tbl:specfit}, {\it bottom left panel}).
\label{fig:hd104237}}
\end{figure}

X-ray emission is detected at the position of HD 104237
(Figure \ref{fig:hd104237}a).
The light curve (Figure \ref{fig:hd104237}b) seems to vary periodically
and rejects a constant model.
The spectrum (Figure \ref{fig:hd104237}c) needs at least 2T components, which
still seem to have small deviation.
\citet{Skinner1996} succeeded in reproducing the SIS spectrum
by 2T models, but we do not succeed in reproducing both
the SIS and GIS spectra simultaneously,
though we do not find any systematic difference between the spectra.
We also tried to fit the spectra by three models,
commonly absorbed 2T model with free elemental abundance,
2T model with different \NH, and commonly absorbed 3T model,
but the reduced $\chi^{2}$ value did not improve.
The large $\chi^{2}$ seems to be caused by weak spectral features
(e.g. emission lines).
In the discussion section, we refer the physical parameters of HD 104237 to the 
commonly absorbed 2T model.

\subsection{IRAS 12496$-$7650}
IRAS 12496$-$7650 is an embedded Herbig Ae star \citep{Hughes1989} and 
is the most luminous far-infrared source in the Chameleon II cloud.
The \ASCA observation is described in \citet{Yamauchi1998}.

\begin{figure}
\plotone{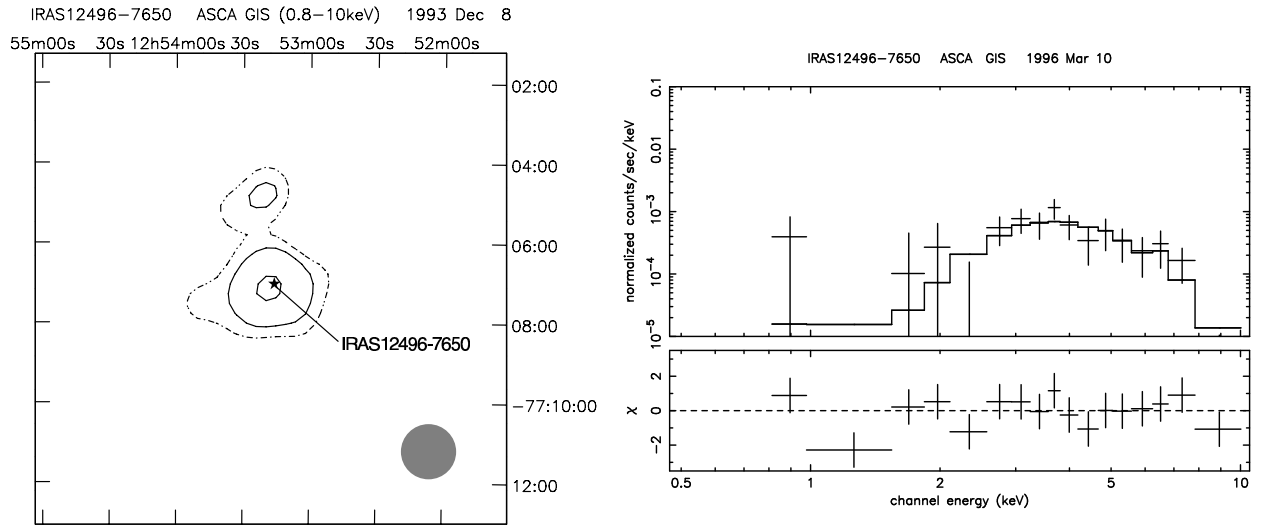}
\caption{a: GIS image of IRAS 12496$-$7650 ({\it left panel}).
b: GIS spectrum of IRAS 12496$-$7650 ({\it right panel}).
\label{fig:iras12496}}
\end{figure}

The GIS image is shown in Figure \ref{fig:iras12496}a.
The X-ray spectrum (Figure \ref{fig:iras12496}b) shows very large 
absorption of about \NH $\sim$ 10$^{23}$ \UNITNH, which
is more than five times of the \NH converted from the \AV \citep{Hughes1991}.
The absorption column is the largest among our sample.

\subsection{HR 5999}
\label{sec:hr59996000}

HR 5999 is a variable A5-7e star with age of 5$\times$10$^{5}$-year-old
located in the center of the Lupus 3 subgroup \citep{Ancker1998,Tjin1989}.
It is thought to have an accretion disk (e.g. \citealt{The1996}),
which might cause photometric and spectral variations.
It also has a close companion at 1\FARCS3 apart (Rossiter 3930, \citealt{Stecklum1995}), 
probably a T-Tauri star (TTS).
The X-ray emission of HR 5999 was detected with \ROSAT \citep{Zinnecker1994}.

\begin{figure}
\plotone{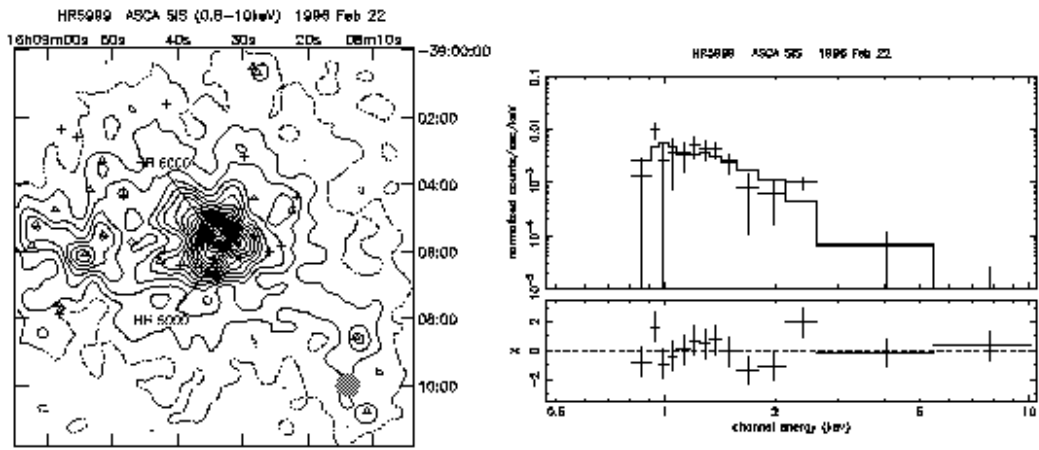}
\caption{a: SIS image of HR 5999 ({\it left panel}).
b: SIS spectrum of HR 5999 ({\it right panel}).
\label{fig:hr59996000}}
\end{figure}

HR 5999 is fainter and hidden in the \PSF wings of the intermediate-mass 
young peculiar star HR 6000.
The peak is only recognized in the SIS image (not clear in Figure \ref{fig:hr59996000}a).
To cancel out contamination from HR 6000, we selected a background region
symmetric with respect to HR 6000.
We do not find any systematic difference in light curve and spectrum between SIS0 and SIS1,
which would mean that contamination from HR 6000 is satisfactorily removed.
Net event counts within a very small ellipse centered at HR5999 (see Table \ref{tbl:reg}) is 
around 50\% of total counts of the source region.
The SIS spectrum of HR 5999 is reproduced by an absorbed 1T thermal model
(Figure \ref{fig:hr59996000}b).
The \NH error range ($<$1.2$\times$10$^{22}$ \UNITNH) includes the \NH converted from \AV ($\sim$0.5$^m$).
The temperature is consistent with the hot component seen in the \ROSAT spectrum (1T model in Table \ref{tbl:specfit}),
and become closer to it by assuming the same \NH as in the \ROSAT analysis
\citep[1T$^F$ model in Table \ref{tbl:specfit},][]{Zinnecker1994}.
The consistency between \NH and \AV and larger \LX than in low-mass stars 
(see discussion \ref{sec:lowmass})
suggests that the detected X-ray emission is from HR 5999.

\subsection{V921 Sco}
\label{sec:v921sco}

V921 Sco (CoD $-$42\DEGREE 11721) is located in the galactic 
plane [($l, b$) = (343.35, $-$0.08)].
Some papers classify it as a supergiant (e.g. \citealt{Hutsemekers1990}).
The distance to V921 Sco has large uncertainty from 200 pc
to 2.6 kpc \citep{Pezzuto1997,Brooke1993}.
We adopt the distance of 500 pc derived from a spectral analysis with the ISO$-$SWS
\citep{Benedettini1998}.

\begin{figure}
\plotone{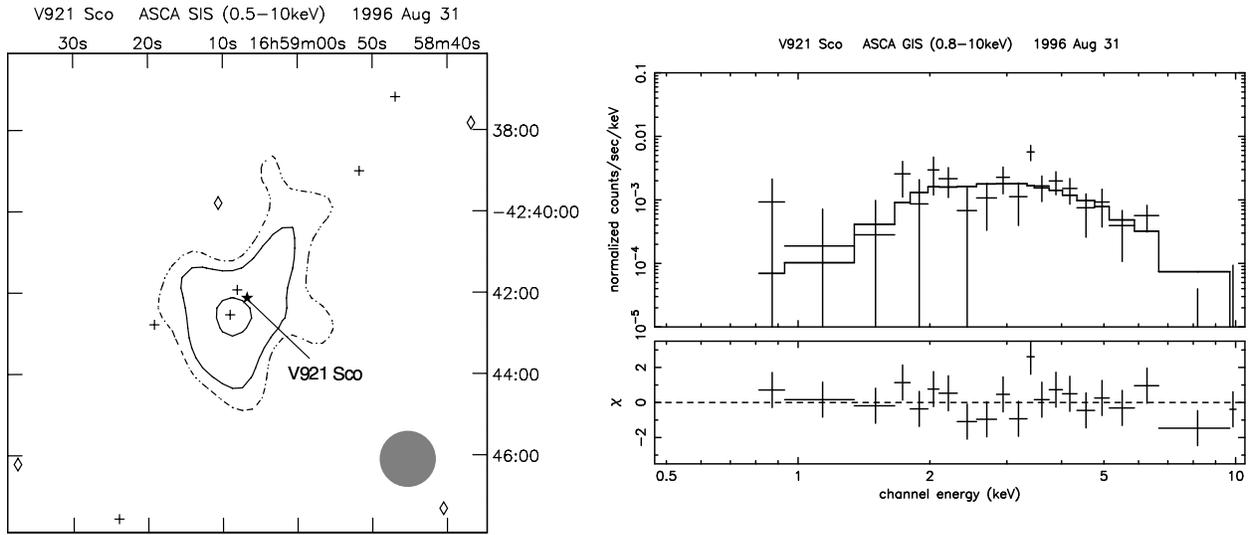}
\caption{a: GIS image of V921 Sco  ({\it left panel}).
b: GIS spectrum of V921 Sco  ({\it right panel}).
\label{fig:v921sco}}
\end{figure}

A weak X-ray source is detected on the GIS image (Figure \ref{fig:v921sco}a).
The source is seen as two marginal humps in the SIS, which might mean that
the X-ray emission is from multiple nearby sources.
Actually V921 Sco, which is in the crowded region on the galactic plane, have
$\sim$50 2MASS sources within 1\ARCMIN though some of them could be nebulae.
We thus show the result of timing and spectral analysis, but do not include the source
for studies in the discussion sections.
The GIS spectrum (Figure \ref{fig:v921sco}b) can be reproduced by an absorbed 1T 
model with strong absorption (\NH $\sim$5$\times$10$^{22}$ \UNITNH).
The \NH is significantly higher than \NH converted from its \AV (\NH $\sim$1.6$\times$10$^{22}$ \UNITNH),
though it is consistent with \NH derived from a millimeter survey 
\citep[$\sim$10$^{23}$ \UNITNH,][]{Henning1998}.

\subsection{MWC 297}
MWC 297 is a highly reddened star with extremely strong Balmer and silicate lines.
\citet{Hillenbrand1992} reported that MWC 297
is a PMS star with the spectral type of O9 at a distance of 450 pc
and age of $\sim$3$\times$10$^{4}$ years.
\citet{Drew1997}, however, suggested it 
to be a B1.5 zero-age main-sequence (ZAMS) star at the distance of $\sim$ 250 pc.
They also argued that MWC 297 is a rapid rotator with $v_{rot}$ sin $i \sim$ 350 \UNITVEL.
The \ASCA observation of MWC 297 is described in \citet{Hamaguchi2000}.
Here, the X-ray luminosity is smaller than that in \citet{Hamaguchi2000}
because we assume a distance of 250 pc.

\begin{figure}
\plotone{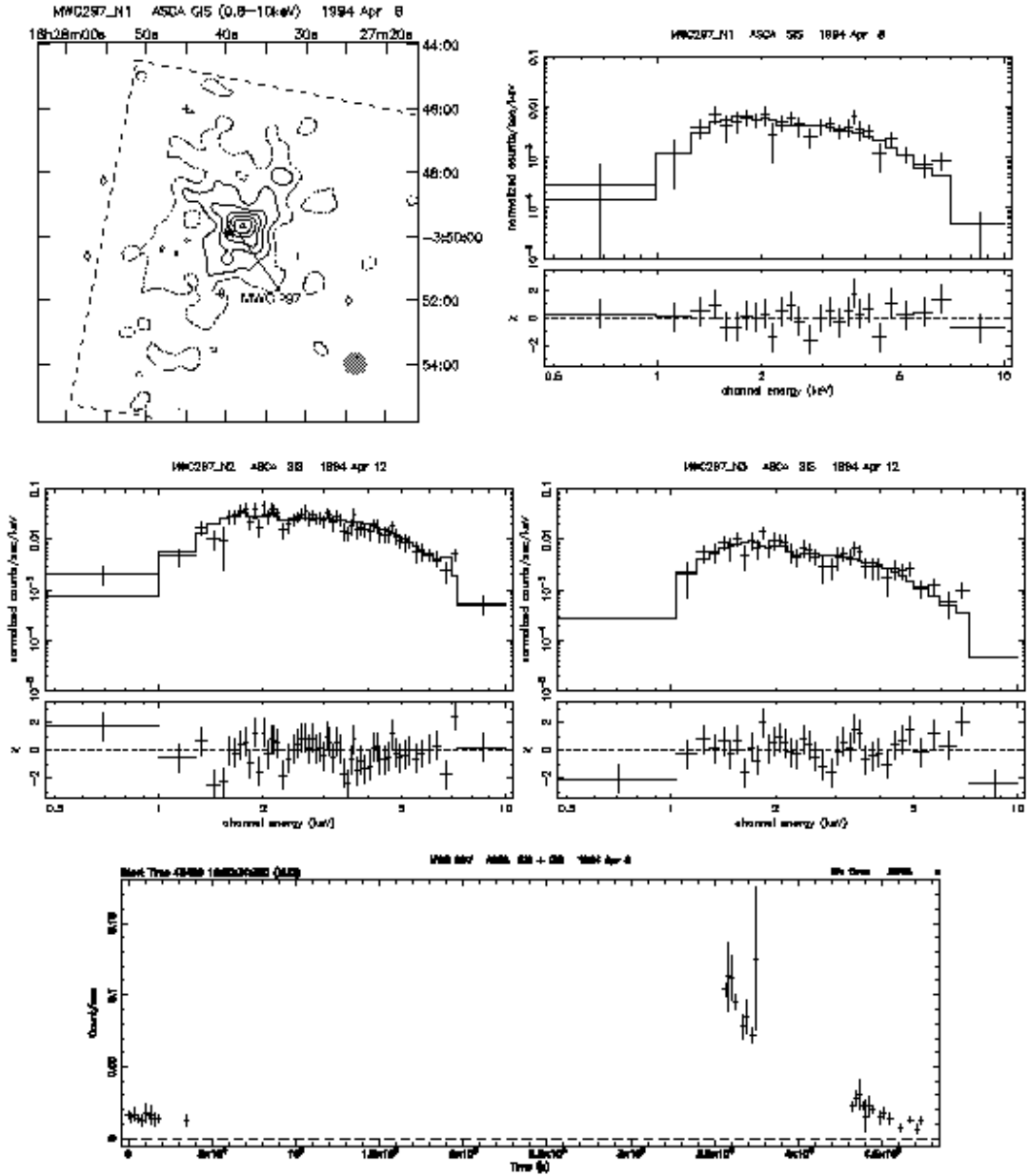}
\caption{a: SIS image of the MWC 297 field in "MWC 297 1" ({\it top left panel}).
The count rate per solid-line is 0.2 \UNITCPPIX .
b: Light curve of MWC~297  ({\it bottom panel}).
c: SIS spectra of MWC 297 in the "MWC297 1, 2 and 3" ({\it top right, middle left, and 
middle right panels, respectively}).
\label{fig:mwc297}}
\end{figure}

A point-like X-ray source is detected in three observations spanning 5 days. 
Figure \ref{fig:mwc297}a shows the SIS image
from the 1st observation (MWC297 1).
We also checked a \ROSAT observation of the MWC 297 field (exposure $\sim$1 ksec), 
but we found no X-ray source in the field.
The combined light curves of the three observations (Figure \ref{fig:mwc297}b) show that
the X-ray flux in the first observation is constant,
but increased by a factor of five at the beginning of the second observation, 
then gradually decreased exponentially to nearly the same flux of the pre-flare level
by the end of the third observation 
with an $e$-folding time $\sim$ 5.6 $\times 10^{4}$ sec.
SIS spectra of all observations are separately 
displayed in Figure \ref{fig:mwc297}c.
The absorption column $\sim$2.4$\times$10$^{22}$ \UNITNH
did not change during the observations.
The plasma temperature in the quiescent phase of 3.4 keV 
increased to 6.5 keV during the flare and then decreased to 3.4 keV.
The flare peak was missed; we only detected X-rays after the on-set of the flare.
Thus, the peak luminosity is
log \LX $\gtrsim$ 32.1 \UNITLUMI, which is the maximum at the beginning of the second observation.
A giant X-ray flare from the early MS star $\lambda$ Eridani (B2e) was detected by \ROSAT 
\citep{Smith1993}. The star would be young and would not be a classical Be star. The flare
detected on MWC297 might have some relation in mechanism to $\lambda$ Eridani.

\subsection{TY CrA / HD 176386}
\label{sec:tycrahd176386}
\label{subsec:tycradr}

TY CrA and HD 176386 belong to the R CrA cloud and
illuminate the reflection nebula NGC 6726/7 \citep{Graham1991}.
TY CrA is a detached Herbig Be eclipsing binary consistency of a 3.16M$_{\odot}$
primary star near ZAMS and a 1.64M$_{\odot}$ secondary star
with an orbital period
of $\sim$2.889 days \citep{Kardopolov1981,Casey1993,Casey1995,Casey1998}.
Both stars are $\sim$3$\times$10$^{6}$-year-old.
The lack of infrared excess indicates absence of an optically thick disk 
\citep{Casey1993}.
It shows non-thermal radio emission \citep{Skinner1993}.
The X-ray emission is detected by \EINSTEIN and \ROSAT.
\citet{Damiani1994} measured the plasma temperature at $\sim$1 keV and
the absorption column density at $\sim$ 5.5$\times$10$^{21}$ \UNITNH.
On the other hand, HD 176386 is a B9.5 star close to the ZAMS with mid-infrared excess
but no H$\alpha$ emission reported
\citep[2.8$\times$10$^{6}$-year-old,][]{Bibo1992,Grady1993,Prusti1994,Siebenmorgen2000}.

\begin{figure}
\plotone{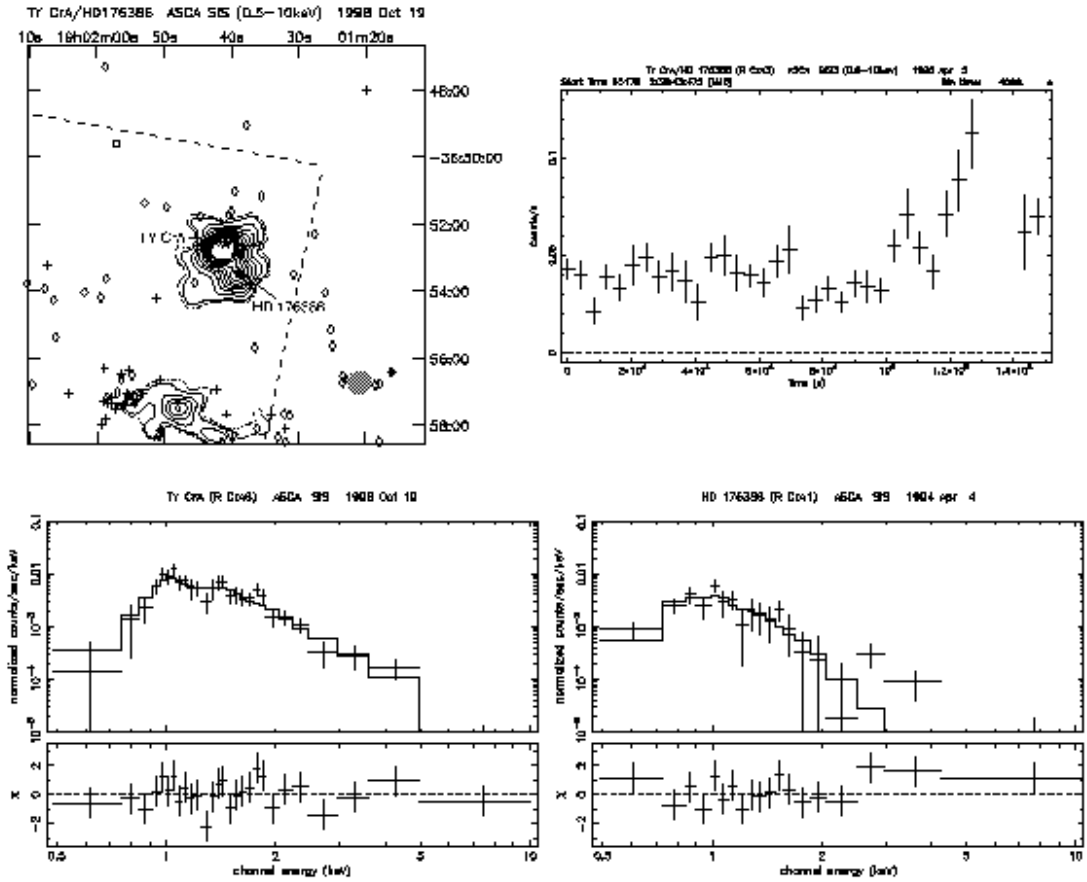}
\caption{a: SIS image of the TY CrA/HD 176386 field in Oct 1998 (R CrA6, {\it top left panel}).
b: GIS light curves of TY CrA + HD 176386 (R CrA3 {\it top right panel}). 
c: SIS spectra of TY CrA (R CrA 6,  {\it bottom left panel}) and HD 176386 (R CrA 1, {\it bottom right panel})
\label{fig:tycra}}
\end{figure}

Both TY CrA and HD 176386 exhibit X-ray emission (Figure \ref{fig:tycra}a).
We focus on the SIS data taken in 3 observations (R CrA 1, 4 and 6)
because the GIS cannot resolve TY CrA and HD 176386, but the SIS marginally does.
We selected a background at the symmetrical region with respect to each source
to cancel contamination from each other.
All light curves are constant in each observation, and the average flux does not change
between observations.
Spectra of both TY CrA and HD 176386 can be fit by an absorbed 1T model
(Figure \ref{fig:tycra}c).
For the three observations, the best-fit column density and plasma temperature 
for TY CrA are 
3$-$6$\times$10$^{21}$ \UNITNH and 1$-$2 keV respectively, which are the
same as the \EINSTEIN result \citep{Damiani1994}.
The fits to the HD 176386 spectrum have two local $\chi^2$ minima, among which smaller 
\NH ($<$2$\times$10$^{21}$ \UNITNH) is consistent with the {\it A$_{\rm V}$}\footnote{\citet{Cardelli1989} suggested \RV to the direction of TY CrA
considerably high compared with normal regions, and
derived larger \AV ($\sim$3.1$^{m}$) than \citet{Hillenbrand1992}(\AV$\sim$1.0$^{m}$).
However, the \AV$-$\NH relation in \citet{Ryter1996} assumes
\RV at 3.1.
We therefore refer to \citet{Hillenbrand1992}.}.
The plasma temperature is around 1$-$2 keV.
On the other hand, at the end of R CrA 3 in which only GIS data is available and hence
X-ray emission from TY CrA and HD 176386 is merged together, 
we see a flare like variation with long rise time ($\sim$10 ksec, see also Table \ref{tbl:flarecurvefit}) as a stellar flare
(Figure \ref{fig:tycra}b).
We define the high flux state HS and the other LS, and 
tentatively fit the spectra by an absorbed 1T model.
Then the best-fit plasma temperature is slightly higher during HS (\KT $\sim$3.0 keV) with 
respect to LS (\KT $\sim$2.6 keV).
Because the rise time-scale is too long for a normal stellar flare, the flare might relate to
binary interaction.
The peak came after 80 ksec of the eclipse of the primary star, equivalent to $\sim$120\DEGREE rotation of the binary system, when we see $\sim$2/3 of the surface of the companion
facing the primary star.
The flare may occur on an active spot of the companion at the root of inter-binary magnetic field.

\subsection{HD 200775}
\label{subsec:hd200775}

HD 200775 is a B2.5e star 
illuminating the reflection nebula NGC 7023 \citep{Ancker1998,Benedettini1998}.
It is in a biconical cavity that has been swept away
by energetic bipolar outflows, but there is no evidence of
ongoing outflow activity (\citealt{Fuente1998}).
Its age is estimated to be $\sim$2$\times$10$^{4}$ years from
the HR diagram,
but it might be 2 $\times$10$^{7}$ year-old in the post main-sequence stage \citep{Ancker1997}.
HD 200775 has a companion star 2\FARCS25 apart \citep[{\it $\Delta$K} = 4.9$^m$,][]{Li1994,Pirzkal1997}.
\citet{Damiani1994} detected strong X-ray emission
with \EINSTEIN (log \LX $\sim$31.9 \UNITLUMI)
and measured \NH $\sim$5$\times$10$^{21}$ \UNITNH and \KT$\sim$0.8 keV.

\begin{figure}
\plotone{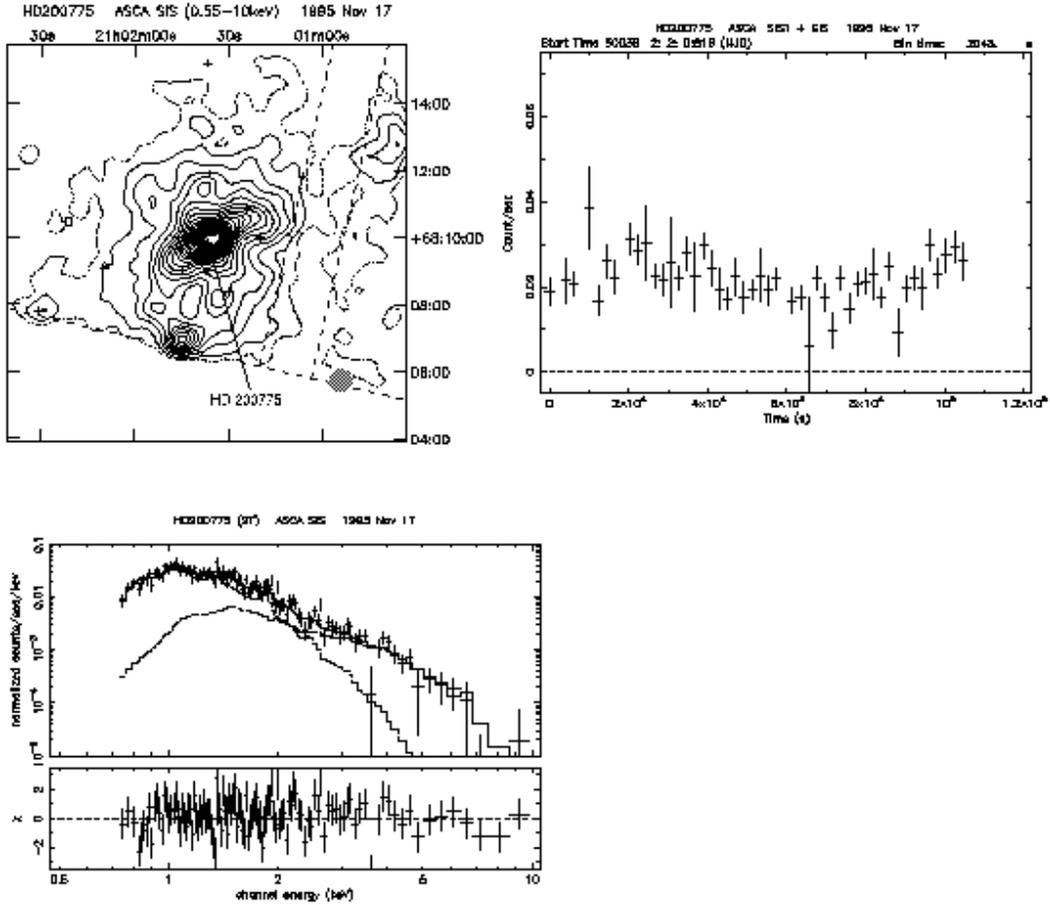}
\caption{a: SIS image of HD 200775 ({\it top left panel}).
b: Light curves (SIS1+GIS) of HD 200775 ({\it top right panel}).
c: SIS spectrum of HD 200775 fit by an absorbed 2T model ({\it bottom left panel}).
\label{fig:hd200775}}
\end{figure}

A point-like source was detected at the position of HD 200775 (Figure \ref{fig:hd200775}a).
The SIS0 light curve has a spike at about 9.3 $\times$10$^{4}$ sec, 
which is probably an instrumental artifact.
The other light curves (SIS1, GIS) show marginal flux increases around 2, 4 and 10$\times$10$^4$ sec,
and the light curve rejects a constant model above 96\% confidence level.
(Figure \ref{fig:hd200775}b).
The spiky event does not have significant counts so that
we use all the SIS and GIS data for spectral fitting (Figure \ref{fig:hd200775}c).
An absorbed 1T model is rejected over 99.9\% confidence.
An absorbed 2T model with common \NH improves 
the reduced $\chi^{2}$ value to 1.15 and is accepted above 90\% confidence.
However, the \NH value ($\sim$9$\times$10$^{21}$ \UNITNH) is two times as large as the 
converted \NH from visual extinction of HD 200775 (\AV$\sim$1.9$^{m}$).
Marginal bumpy features are present in the 2$-$4 keV band. 
We thus tried these two models: (i) 2T thermal model with a free abundances, 
(ii) 1T thermal and hard power-law model.
In the model (i), the metal abundance is slightly larger (0.39 solar) than the fixed 
abundance (0.3 solar), but the model fit does not improve.
On the other hand, the best-fit \NH of the model (ii) is $\sim$4$\times$10$^{21}$\UNITNH, 
which is almost equal to the \NH converted from \AV though we do not have physical interpretation to explain
the power-law component.

\subsection{MWC 1080}

\label{sec:mwc1080}
MWC 1080 is an early A type star with \AV $\sim$4.4$^{m}$\citep{Yoshida1992}.
The distance is 2.2--2.5 kpc from UVBR photometry of
neighboring MS stars \citep{Grankin1992} and kinematic distance
of $^{13}$CO molecular line \citep{Canto1984}.
It forms a multiple system of a \HAEBE primary, 
a plausible \HAEBE companion (0\FARCS7) and another companion (4\FARCS7),
with eclipse events at $\sim$2.887 days \citep{Grankin1992,Leinert1997,Pirzkal1997}.
It has CO molecular outflow and ionizing winds up to 1000 \UNITVEL
\citep{Canto1984,Yoshida1992,Leinert1997,Benedettini1998}.
The X-ray emission was detected with \ROSAT \citep[log \LX $\sim$32 \UNITLUMI]{Zinnecker1994}.

\begin{figure}
\plotone{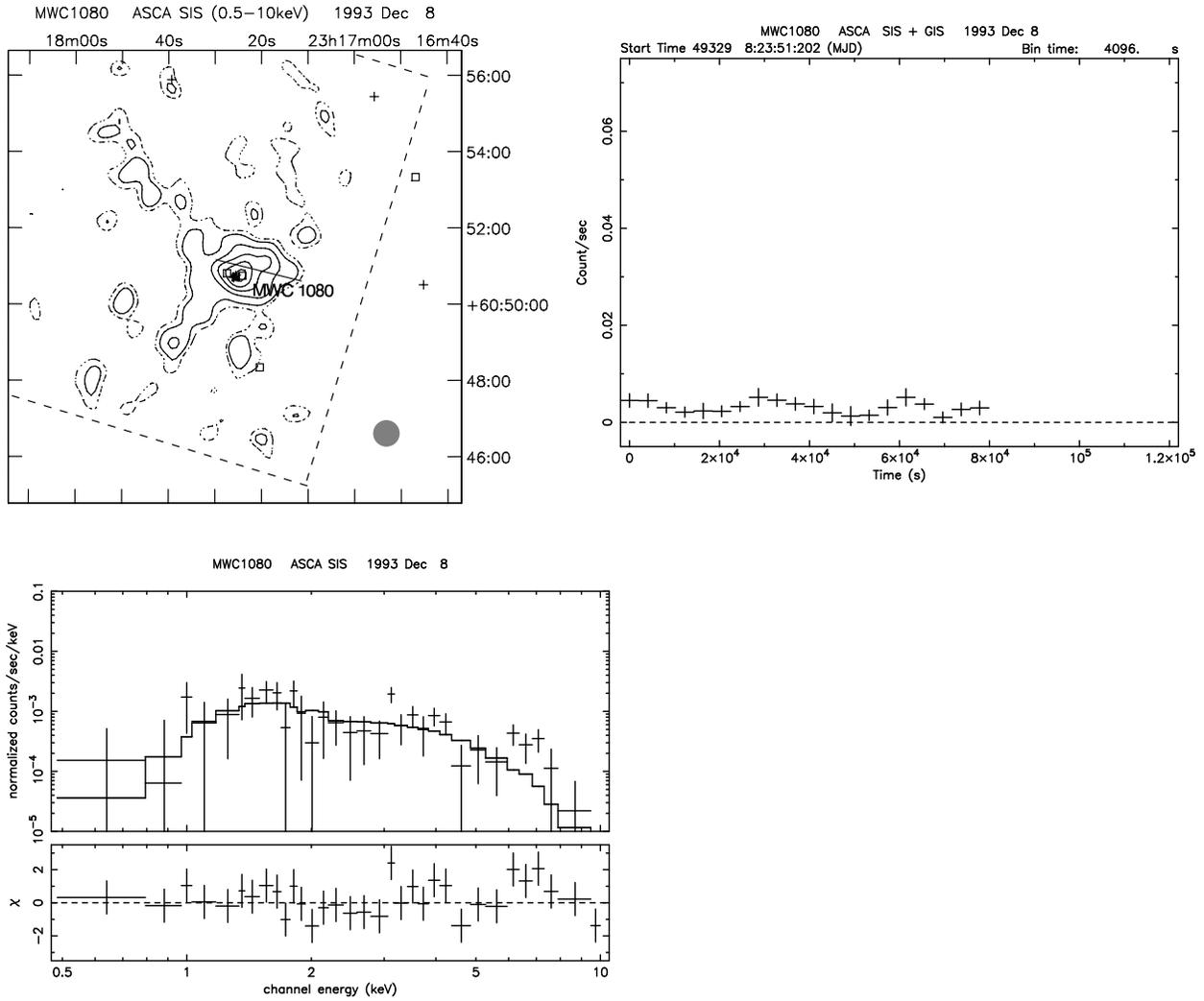}
\caption{a: SIS image of MWC 1080 ({\it top left panel}).
The count rate per solid-line is 0.2 \UNITCPPIX.
b: Light curves of MWC 1080  ({\it top right panel}).
c: SIS spectrum of MWC 1080  ({\it bottom left panel}).
\label{fig:mwc1080}}
\end{figure}

An X-ray source was detected near MWC 1080 (Figure \ref{fig:mwc1080}a).
The light curves (Figure \ref{fig:mwc1080}b) seem to oscillate periodically
by $\sim$40 ksec, but this is not statistically significant.
The orbital phase during the observation is $\phi =$ 0.39 to 0.74 according to
the ephemeris by \citet{Grankin1992}.
The secondary eclipse occurs at 3$\times$10$^{4}$ sec, but it is not observed in the 
X-ray light curve.
The light curves do not show evidence that binarity affects the observed X-ray emission of
MWC 1080.
The spectrum (Figure \ref{fig:mwc1080}c) can be reproduced by
an absorbed 1T model.
The temperature is extremely high (\KT $>$3.8 keV) and 
the absorption column density (\NH $\sim$10$^{22}$ \UNITNH) is consistent with \AV.
The spectrum has an excess around 6--7 keV.
When we include a Gaussian model with $\sigma$= 0.0,
the center energy is 6.4 (6.2--6.6) keV, which corresponds to the iron fluorescent line.
On the other hand, an absorbed 1T model with the free abundance parameter
has abundance of 0.8 solar, which is mainly determined from Helium-like Fe-K line
emission.
The residual still seems to have systematic patterns though the model is accepted
$>$90\% confidence. A commonly absorbed 2T model or soft 1T + power-law model compensate
the pattern slightly better, but the best-fit \NH ($\sim$5$\times$10$^{22}$ \UNITNH) is large and luminosity of the best-fit model is 
unrealistically high for stellar X-ray emission ($>$10$^{36}$\UNITLUMI).

\subsection{EC 95}
\citet{Preibisch1998} first reported constant X-ray emission from EC 95.
From near-infrared spectroscopy,
\citet{Preibisch1999} determined EC 95 as an extremely young 
intermediate-mass star
(Age $\sim$2$\times$10$^{5}$ years; $M \sim$4 \UNITSOLARMASS), 
a so called proto-\HAEBE.
The very high visual extinction (\AV $\sim$25$-$35$^{m}$)
made them suspect that the source has a large intrinsic X-ray luminosity
of about log \LX $\sim$33 \UNITLUMI.
Radio observations have detected a point-like source, SH 68-2,
within the error circle of EC 95 \citep{Rodriguez1980,Snell1986,Smith1999},
showing gyrosynchrotron emission suggesting presence of a magnetosphere.

\begin{figure}
\plotone{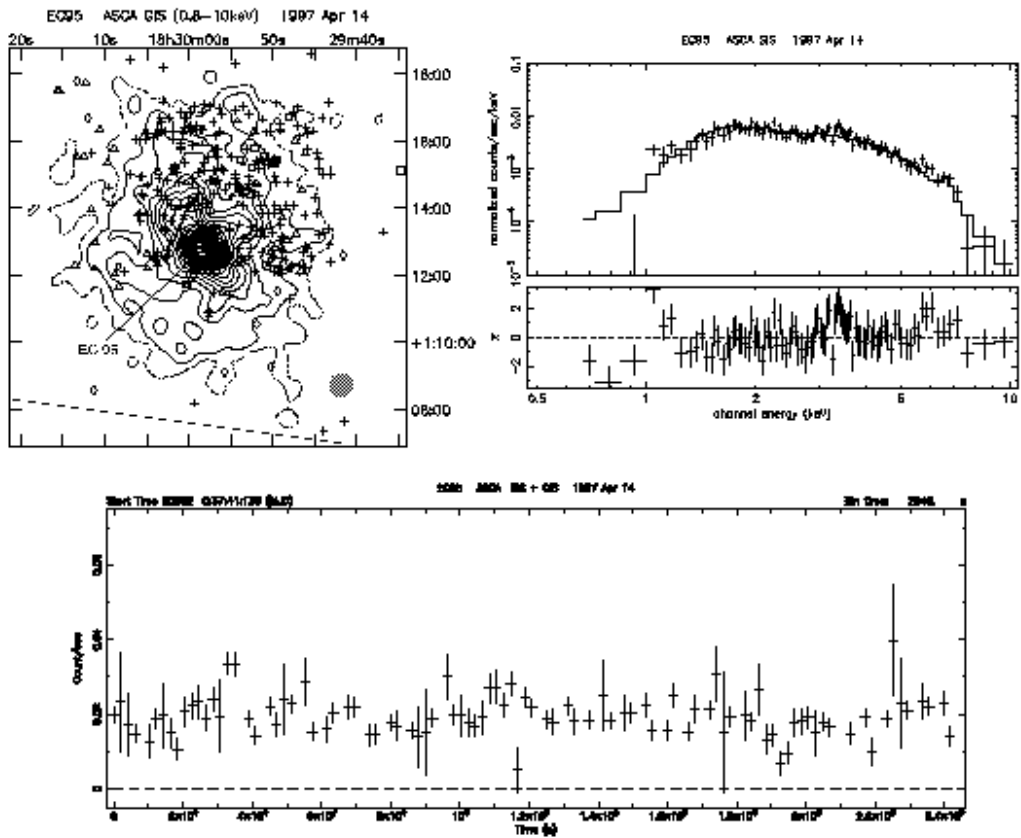}
\caption{a: SIS image of the EC 95 field ({\it top left panel}).
b: Light curves of EC 95  ({\it bottom panel}).
c: SIS spectrum of EC 95  ({\it top right panel}).
\label{fig:ec95}}
\end{figure}

A point-like source is detected within the error circle of EC 95 (Figure \ref{fig:ec95}a).
The nearest \ROSAT X-ray source is 1\FARCM5 apart 
\citep{Preibisch1998}.
\ASCA would see the same source.
The light curves show flare-like small fluctuations and  
reject a constant model (Figure \ref{fig:ec95}b).
Spectra of the SIS and GIS (Figure \ref{fig:ec95}c for the SIS) are inconsistent 
in absolute flux.
The SIS flux calibration is less reliable due to CCD degradation.
We independently varied the model normalizations, and
find an acceptable result.
The flux is derived from the GIS data.
The plasma temperature is quite high (\KT $\sim$3.9 keV). 
\NH is large but about one fourth of the \AV ($\sim$36$^{m}$) converted \NH.
The luminosity (log \LX $\sim$31.7 \UNITLUMI) is therefore about an order of mag 
smaller than the \ROSAT estimate.

\section{X-ray Emission Mechanism}

\subsection{Could the X-ray Emission Come from a Low-Mass Companion?}
\label{sec:lowmass}

More than 85\% of \HAEBEs are visible or spectroscopic binaries
\citep{Pirzkal1997}, whose 
companions could be low-mass stars, i.e. TTSs or protostars.
Low-mass stars are known to emit strong X-rays compared with their
faint optical and IR emission (\LX/\Lbol up to 10$^{-3}$).
We thus evaluate as whether the X-ray emission observed with \ASCA
could come from low-mass stars.
\citet{Stelzer2001} reports that, in the Taurus Auriga region, 
CTTSs and WTTSs with log \LX $\gtrsim$30 \UNITLUMI are around 5\% and 25\%, respectively.
Multiplying the binary ratio,
around 4\% and 20\% of the X-ray emission with log \LX $\gtrsim$30 \UNITLUMI
could be really from low-mass companions.
\ASCA detected sources have log \LX $\gtrsim$ 30 \UNITLUMI and the detection
rate is $\sim$30\%.
Contribution of the companion could be non-negligible if the companions follow
the luminosity function of WTTSs.
However, considering that the primary star and its companion tend to have similar ages (e.g. see \citet{Casey1998} for 
TY CrA), 
CTTSs with the similar ages to \HAEBEs ($\sim$10$^{6}$-year-old) might be more appropriate for this estimate, then the
contribution of the companion in our sample would be negligible.
While there are almost no TTSs above 10$^{31}$ \UNITLUMI, therefore the
X-ray emission would be responsible for the primary (i.e. early-type) star.

\subsection{X-ray Properties}
\label{sec:xrayhaebemechansim}

The X-ray properties of \HAEBEs and their relation to the stellar parameters
may provide further insight into the origin of the X-ray emission.
In this section, we examine the relation of \LX to other stellar parameters such as \Lbol, \Teff 
and v$_{rot}$ sin $i$,
the X-ray time variability and the plasma temperature. We then compare the properties with those of
low-mass and high-mass stars.
In the discussion of X-ray luminosity,
we included \ROSAT samples of \HAEBEs \citep{Zinnecker1994}.
Their \LX in the \ASCA band (0.5--10 keV) is estimated by 
assuming an absorbed thin-thermal model with \KT = 2 keV, abundance = 0.3\UNITSOLARABUND 
and \NH converted from \AV \citep{Ryter1996}.
The errors are put between half and factor of two of their \LX to allow for uncertainty of
\KT $\sim$0.5$-$4 keV.
The photon statistical errors are basically much smaller than this error range.
Whereas, we also include EC 95 and SSV 63E+W as candidates of proto-\HAEBEs.
SSV 63E+W is either a Class I protostar candidate SSV 63E or W in the Orion cloud.
\citet{Zealey1992} mention that SSV 63E may have mass 
1.5\UNITSOLARMASS $<M<$5 \UNITSOLARMASS assuming age less than 10$^{6}$-years
at a distance of 460 pc. 
The X-ray emission is reported by \citet{Ozawa1999}.
On the other hand, we did not include V921 Sco in the sample.

\begin{figure}
\plotone{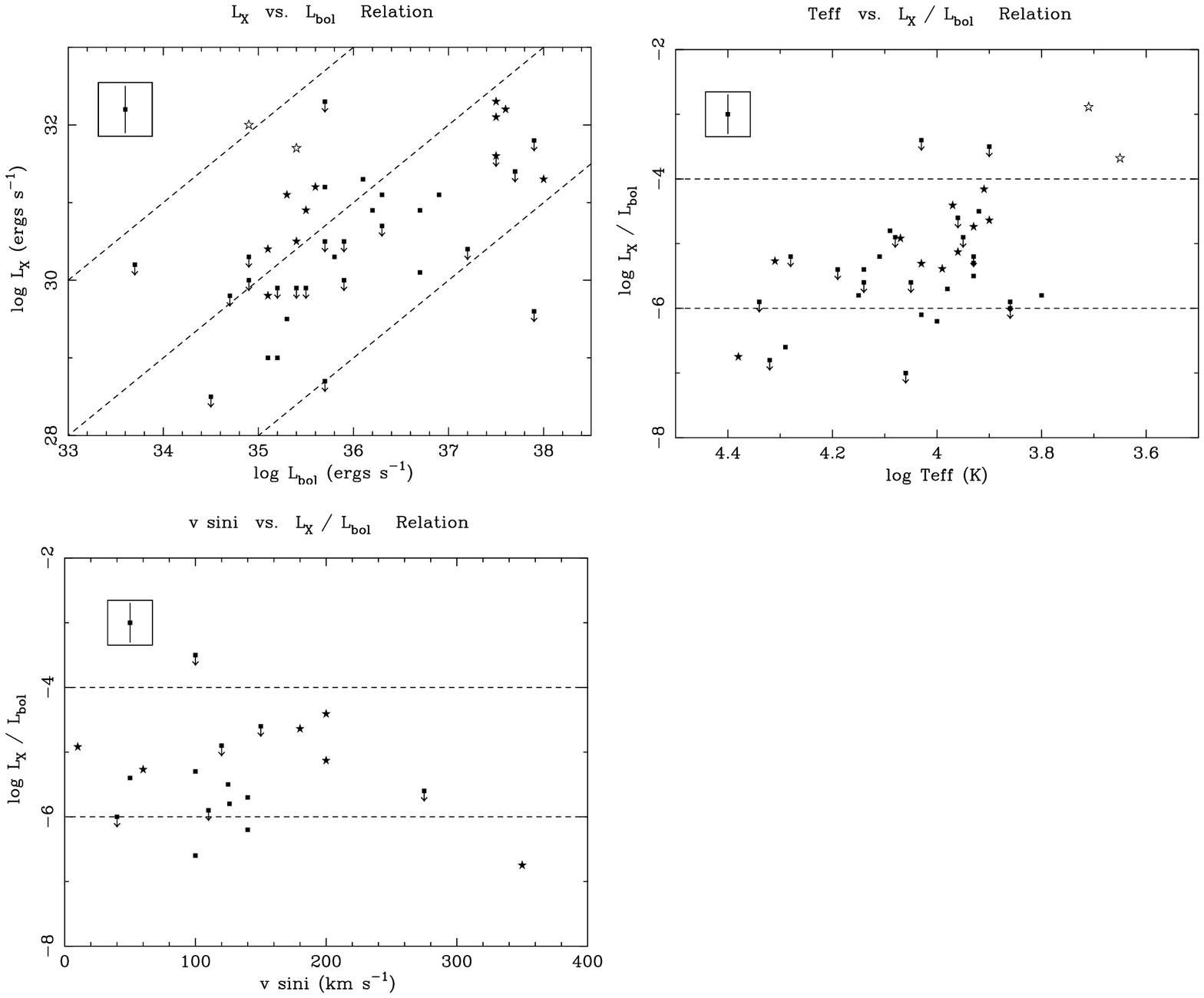}
\caption{a: Log \LX $vs.$ log \Lbol relation ({\it top left panel}),
b: log \LX/\Lbol $vs.$  log \Teff relation ({\it top right panel}),
c: log \LX/\Lbol $vs.$ $v_{rot}$ sin $i$ ({\it bottom left panel}).
Filled and open stars represent HAeBes and proto-HAeBes of \ASCA samples, respectively.
Filled squares represent HAeBes of \ROSAT samples.
Points with arrows represent upper-limit.
\LX of \ROSAT is converted to the \ASCA band assuming \KT = 2 keV, abundance = 0.3\UNITSOLARABUND and \NH converted from \AV \citep{Ryter1996}.
The plot in the upper left box represents typical error ranges of \ROSAT samples if \KT 
changes between 0.5--4 keV.
Barred lines in the figure (a) show log \LX/\Lbol = $-$3, $-$5 and $-$7 from the top, respectively.
\label{fig:lxlbolrelation}
\label{fig:tefflxlbolrelation}}
\end{figure}

Figure \ref{fig:lxlbolrelation}a shows the \LX $vs.$ \Lbol diagram.
In our \HAEBE samples ranging between 33 $<$log \Lbol \UNITLUMI $<$38,
the log \LX/\Lbol ratio is between $-$4 and $-$7.
The ratio is above the range of massive MS stars ($-$6 $\sim-$8), and rather
close to, but slightly different from, that of low-mass stars (up to $-$3) \citep[e.g.][]{Gagne1995}.
The \LX/\Lbol ratio of proto-\HAEBE is close to $-$3.
This might suggest that X-ray activity is enhanced in the very youngest stars
(see also discussion in the section \ref{subsec:nhktrel}).
Figure \ref{fig:tefflxlbolrelation}b shows the dependence of log \LX/\Lbol on log \Teff.
The log \LX/\Lbol increases as log \Teff decreases, which resembles 
the trend of X-ray sources in the Orion cloud \citep[Fig. 8 in][]{Gagne1995}.
Figure \ref{fig:lxlbolrelation}c shows the log \LX/\Lbol dependence on $v_{rot}$ sin $i$
\citep{Grady1993,Boehm1995,Drew1997,Casey1998,Ancker1998,Corporon1999}.
The log \LX/\Lbol of low-mass MS stars saturates at $\sim-$3 above $v_{rot}$ sin $i$ $>$ 15 \UNITVEL \citep[][]{Stauffer1994}, which represents a saturated dynamo on solar-like low-mass stars.
Log \LX/\Lbol for \HAEBEs is 
rather small for high $v_{rot}$ sin $i$, and does not seem to saturate at any specific velocity.
The activity might not be generated by a magnetic dynamo 
as in solar-like low-mass stars.

\begin{figure}
\plotone{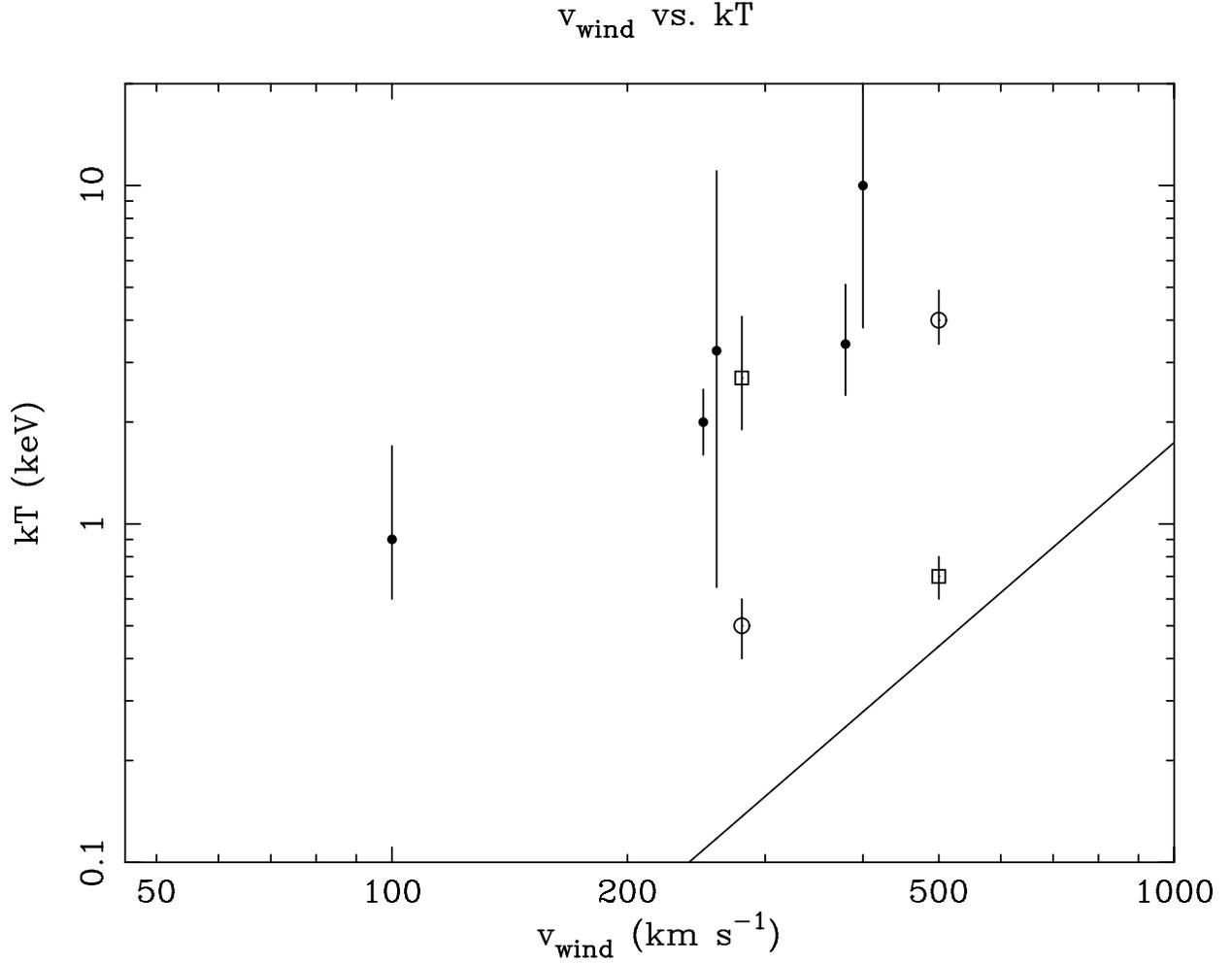}
\caption{\KT $vs.$ terminal velocity of the stellar wind ($v_{wind}$).
Dots are 1T sources.
Open circles and squares are dominant (i.e. large \LX) and minor (i.e. small \LX) components of 2T sources, respectively.
The solid line denotes the plasma temperature when whole kinetic energy 
converts to the thermal energy
(i.e. 3$kT$=$m_{p}v_{wind}^{2}/2$, $m_{p}$: proton mass).
\label{fig:lkinlxrelation}
\label{fig:ktvrelation}}
\end{figure}

The plasma temperature of \HAEBEs (\KT $\sim$2 keV) 
is significantly higher than the typical temperature of high-mass MS stars,
which is driven by high speed stellar winds (\KT $\sim$1 keV, $v_{wind} \sim$1000$-$3000 \UNITVEL).
We calculate the maximum plasma temperature produced by \HAEBE stellar winds 
\citep{Nisini1995,Benedettini1998}
assuming that all wind kinetic energy is thermalized (Figure \ref{fig:ktvrelation}).
Then the plasma temperature observed with \ASCA is above the maximum temperature which can be
made by \HAEBE winds.
While, X-ray emission from $\sim$2 keV plasma is usually seen on TTSs and protostars
with magnetic activity.

\label{subsec:promtimevar}
MWC 297, VY Mon and TY CrA exhibited prominent variability (see Table \ref{tbl:flarecurvefit}).
MWC 297 and TY CrA had strong flux increase from the pre-flare level
and exponential decay seen on low-mass stellar flares, while
VY Mon had rather steady decay.
Variations of MWC 297 and TY CrA strongly indicate a flare magnetic activity
although they could be on their companions.
Their decay timescales derived by exponential model fits
(10$-$60 ksec)
are longer than those of low-mass MS and TTS flares \citep[2$-$7 ksec,][]{Stelzer2000} 
and as long as the flare decay of protostars \citep[10$-$30 ksec,][]{Tsuboi2000}.
X-ray flares on \HAEBEs might be similar to those on protostars.

Plasma temperatures and variability are similar to those of low-mass stars.
This may suggest that X-ray emission from \HAEBEs originates from magnetic activity.
We have enough reason to believe it because optical observations suggest that
some HAes have azimuthal magnetic fields
(AB Aur: MgII 2795\AA, \citealt{Praderie1986}; HeI 5876\AA, \citealt{Boehm1996}).
On the other hand, the \LX/\Lbol ratio and its dependence on stellar rotational velocity is different
from those of low-mass stars, which
may mean that magnetic activity is not by a magnetic dynamo.
Actually, the stellar evolutional model of intermediate-mass stars \citep{Palla1990} predicts
the absence of a surface convection zone necessary for a solar type dynamo.
Another type of magnetic activity might be required for \HAEBEs.

\subsection{Evolution of Stellar X-ray Activity}

\label{subsec:nhktrel}
\begin{figure}
\plotone{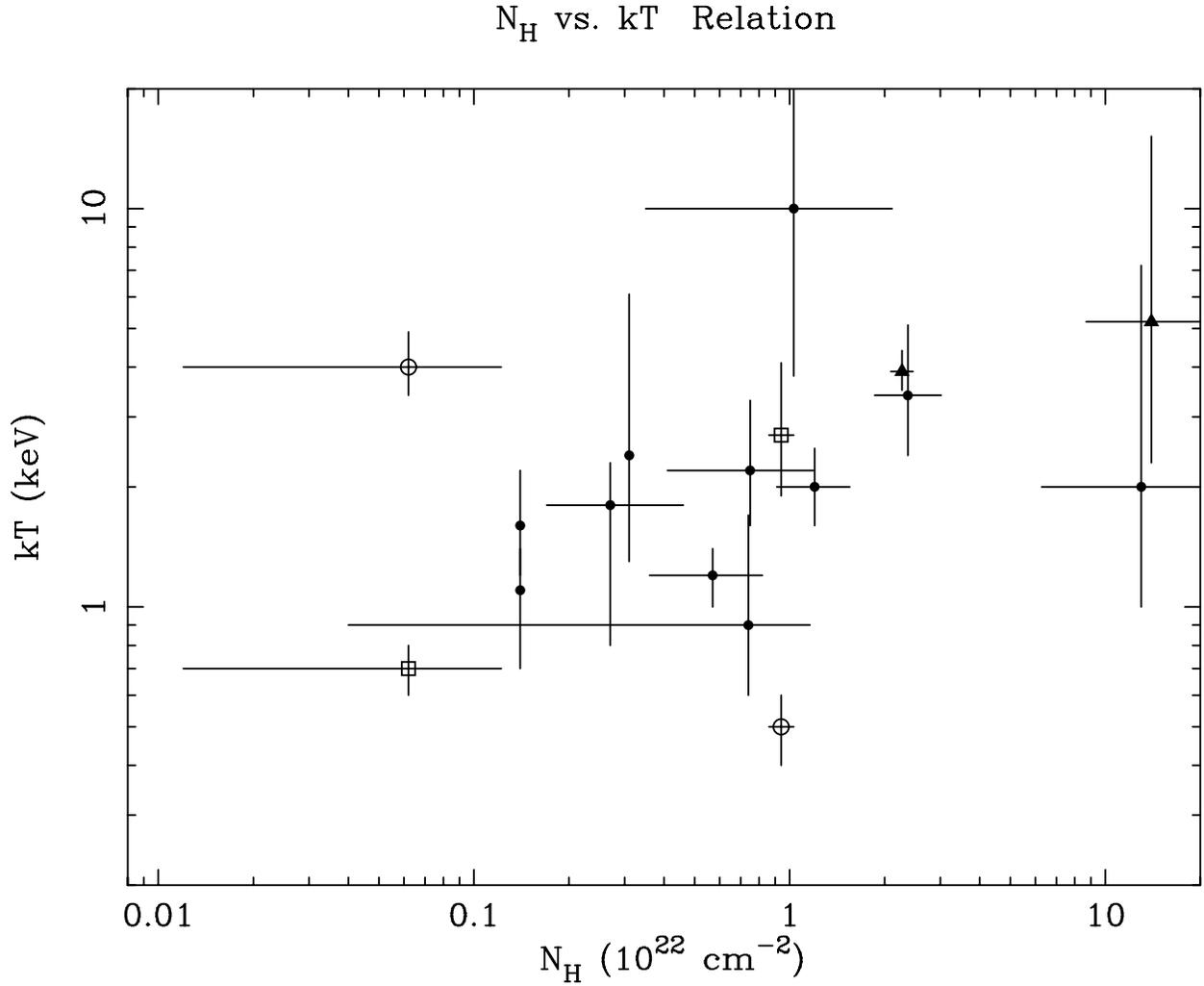}
\caption{\KT dependence on \NH.
Dots, open circles and squares are the same as in Figure \ref{fig:lkinlxrelation}.
Proto-HAeBes (1T) are shown as filled triangles.
\NH less than 10$^{21}$ \UNITNH is referred to the \AV-converted \NH.
\label{fig:nhavktrelation}}
\end{figure}

Circumstellar gases gradually dissipate through mass accretion, outflows and stellar winds.
The hydrogen column density along the line of sight therefore decreases with age,
although it might also depends on stellar mass, disk inclination and so on.
We thus regard \NH as a rough indicator of stellar age.
We then see \KT rises from 2 to 5 keV as \NH increases  (Figure \ref{fig:nhavktrelation}).
Points in the high \NH low \KT area are missing simply because of the weak sensitivity
to soft X-ray emission from large \NH sources, but 
plasma temperature tends to increase with \NH.
Younger \HAEBEs tend to have hotter temperature.

\begin{figure}
\plotone{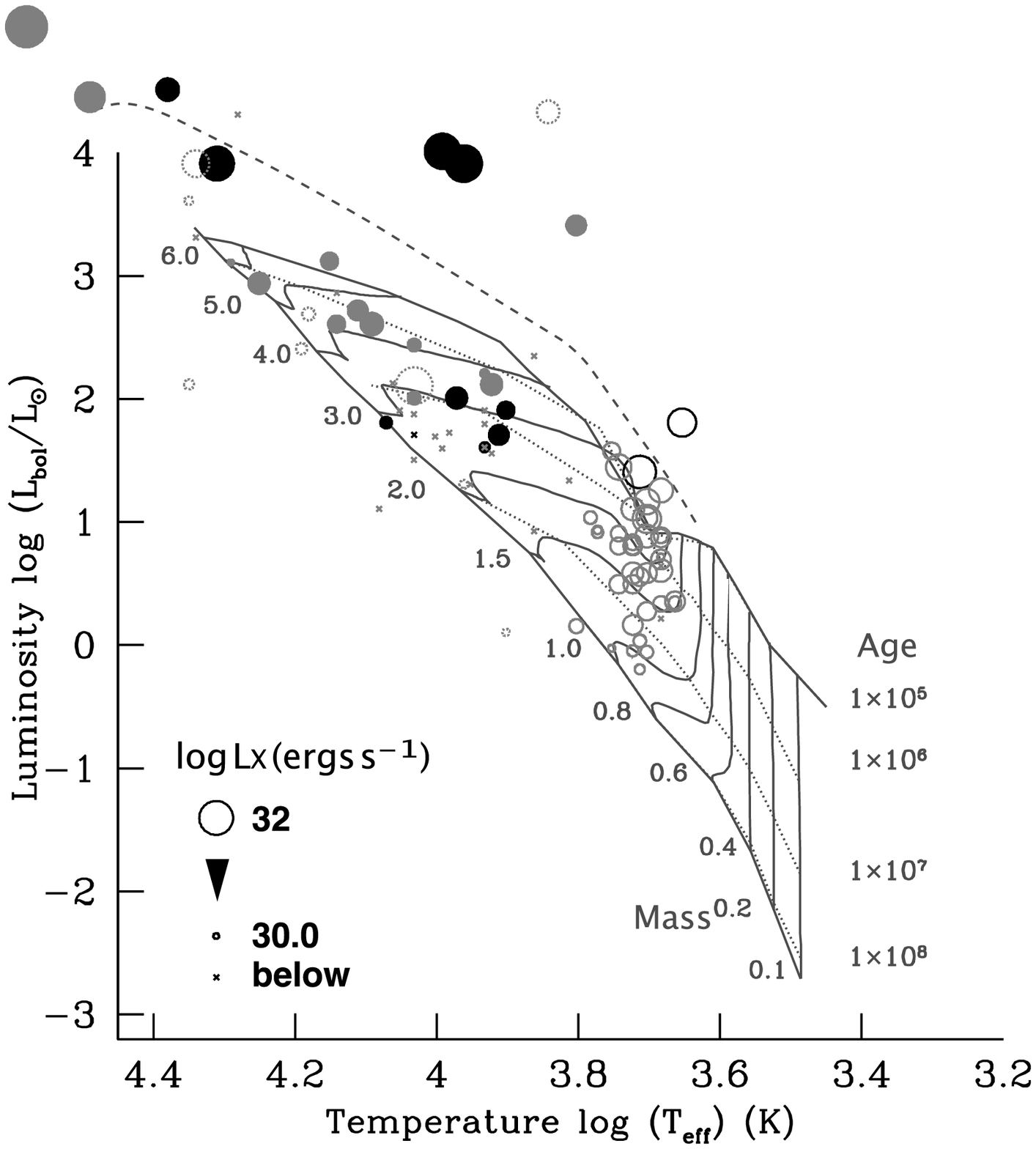}
\caption{X-ray luminosity of young intermediate-mass stars in the HR diagram.
Circle size is proportional to (log \LX \UNITLUMI $-$ 30) and 
crosses are log \LX $<$30 \UNITLUMI (both detected and upper limit sources).
Luminosity upper limit on log \LX $\geq$30 \UNITLUMI is drawn by dotted circles.
Black and grey are \ASCA and \ROSAT samples, respectively.
\HAEBEs are drawn by filled circles and proto-HAeBe candidates
(EC 95 and SVS 63E+W) are two black open circles.
Grey open circles and crosses below log T$_{\rm eff} \sim$3.8 K are sources in the Orion cloud \citep{Gagne1995} (Undetected sources with upper limit above 10$^{30}$ \UNITLUMI are not included).
The diagram is overlaid on a stellar evolutional track modeled by \citet{Palla1993}.
A broken line means the birth line when accretion rate 
is \Mdot = 10$^{-4}$ \UNITSOLARMASS yr$^{-1}$.
\label{fig:graphhrevolution0500all}}
\end{figure}

\label{subsec:evolutionhrdiag}
\LX does not have clear correlation with \NH.
We thus make a bubble chart of \LX in the HR diagram
(Figure \ref{fig:graphhrevolution0500all}).
Sources in the upper right region in the diagram have
log \LX $\sim$32 \UNITLUMI and gradually decrease as they come close to the MS branch.
A notable feature is the presence of an {\it X-ray inactive region} in the HR diagram:
3.8 $<$ log \Teff (K) $<$ 4.1 and 
1.3 $<$ log \Lbol (\UNITSOLARLUMI) $<$ 1.9, which corresponds to
the age older than 10$^{6}$-year-old.  Almost no source in this region
has log \LX (\UNITLUMI) $>$30.
This {\it X-ray inactive region} may indicate that X-ray activity of HAeBes terminates
after the  age of $\sim10^6$-year-old.
\citet{Damiani1994} found a correlation between \LX and spectral type or \Lbol,
but \citet{Zinnecker1994} did not.
We now see that this difference is simply due to selection effects.
\citet{Damiani1994} collected \HAEBE samples close to the MS so that
they found a similar X-ray characteristics to MS stars.
\citet{Zinnecker1994} examined young A-type samples such as V380 Ori and HR 5999.
Their X-ray luminosity is comparable to early B-type stars and thus has no dependence
on stellar spectral type.

\label{subsec:stellardisk}
The above discussion suggests that the X-ray activity decays with age 
and falls below log \LX (\UNITLUMI) $\sim$30 at $\sim$10$^{6}$-years-old.
This age coincides with the end phase of outflow activity \citep{Fuente1998}.
We therefore hypothesize that X-ray activity of \HAEBEs is triggered by reconnection 
of magnetic fields linking a star and its circumstellar disk 
(a star and disk dynamo activity) as it was claimed
for the X-ray emission mechanism of low-mass protostars
\citep[e.g.][]{Koyama1996,Hayashi1996,Montmerle2000}.
This model reconciles the lack of a magnetic dynamo on the stellar surface of \HAEBEs.
To test the hypothesis, we correlate \LX of \HAEBEs with outflow activity
by referring to tables in \citet{Maheswar2002} and \citet{Lorenzetti1999}
and several individual papers (Table \ref{tbl:outflow}).
Then, sources with log \LX $>$30 \UNITLUMI tend to have outflow activity, which
might suggest a link between X-ray and outflow activity.
Identification of outflow sources especially in crowded regions (e.g. EC 95, R CrA, T CrA)
seems to have uncertainty, while the outflow from HD 200775 could be made by stellar winds \citep{Fuente1998}. More careful study would be required.

%% A problem lies in that HAes older than 10$^{6}$-year-old, 
%% which do not show strong X-ray activity,
%% still have thick circumstellar disks (e.g. AB Aur, HD 97048).
%% The activity might relate to the inner most structure, which is unresolved
%% even with the radio observations.

Our results suggest that X-ray emission from young \HAEBEs originates in magnetic
activity.
This means that magnetic activity plays an important role for both low-mass and 
intermediate-mass PMSs.
We here propose a unified scenario for the  X-ray activity of PMSs.
In this scenario, stellar X-ray emission begins in the protostar phase
when stars are deeply embedded in molecular cloud cores.
In this phase, stars possess fossil magnetic fields  from
their parent clouds and exhibit violent magnetic activity linking a star and its disk,
showing occasional X-ray flares.
A part of the infalling gases goes out from the system with large amounts of
angular momentum, which is later seen as molecular outflows or optical jets.
After the star-disk dynamo activity disappears in $\sim$10$^{6}$-year-old, low-mass stars still
continue magnetic activity similar to a solar type dynamo.
Stars between A - F5 lack any mechanism to produce hot plasma.
X-ray activity ends for these stars, and the {\it X-ray inactive region} is seen in the HR diagram.

\section{Relation of X-ray Activity to MS OB Stars}
\label{sec:evolveobstar}
\begin{figure}
\plotone{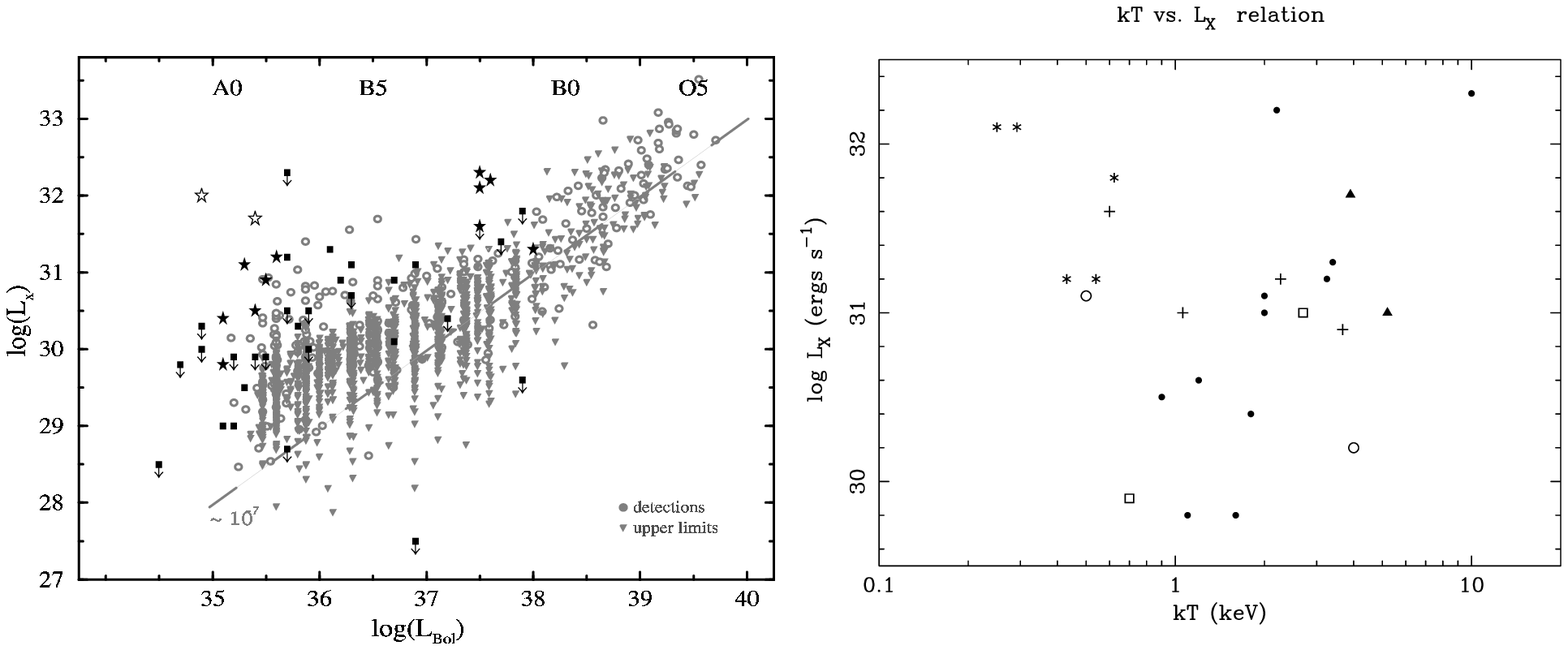}
\caption{a: Log \LX $vs.$ log \Lbol relation of HAeBes {\it (black)}
overlaid on those of MSs ({\it gray}, figure 4 in \citet{Berghoefer1997}) ({\it left panel}).
Marks of \HAEBEs are the same as those in Figure \ref{fig:lxlbolrelation}a.
b: Log \LX $vs.$ \KT relation of massive MS stars and \HAEBEs  ({\it right panel}).
Asterisks are cool components of 2T sources or 1T sources of massive MS stars.
Crosses are hard X-ray tails of massive MS stars.
Other marks for \HAEBEs and proto-\HAEBEs are the same as in Figure \ref{fig:nhavktrelation}.
\label{fig:bergh1997rassob}
\label{fig:ktlxobstarhaebe}}
\end{figure}

Figure  \ref{fig:graphhrevolution0500all} does not show any clear decrease of X-ray activity 
on B-type stars (log T$_{eff} \gtrsim$ 4 (K)),
but the \LX-\Lbol ratio of Herbig Be stars is larger than MS B-type stars
(see Figure \ref{fig:tefflxlbolrelation}b), which suggests some transition of X-ray activity
close to the MS stage.
We superimpose our \HAEBE data on the \LX-\Lbol graph of OB stars in the sky
using \ROSAT all-sky survey data \citep[Figure \ref{fig:bergh1997rassob}a, the relation of OB stars referred to][]{Berghoefer1997}.
Then, most of the HAeBe samples are above the higher end of OB MS stars.
This would mean some transition of X-ray activity between \HAEBE and MS phases.
B-type stars are not generally thought to have accretion disks.
The result may strengthen the star-disk dynamo hypothesis.
On the other hand, we do not see any clear gap between MS and \HAEBE plots.
\citet{Berghoefer1997} suggests that the \LX/\Lbol relation of early-type stars (log \LX/\Lbol $\sim-$7), which is characteristic of a stellar wind origin, scatters later than B1$-$1.5 stars.
This might imply that X-ray activity of \HAEBEs holds in the early stage of MS B-type stars.

Though no relation to the stellar evolution is claimed between O stars and \HAEBEs,
comparison of their X-ray properties may help to understand X-ray activity.
We thus made a \KT and \LX graph (Figure \ref{fig:ktlxobstarhaebe}b) of both types of stars
(O stars$-$ $\delta$ Ori, $\lambda$ Ori; \citealt{Corcoran1994}: $\tau$ Sco;
\citealt{Cohen1997}: $\zeta$ Oph, $\zeta$ Ori; \citealt{Kitamoto2000}).
\LX for O stars are calculated from \EM and \KT from their references, using XSPEC 9.0.
Some O stars have hard X-ray tail \citep[e.g.][]{Corcoran1993,Kitamoto1996},
which is separately shown in the figure.
The hard tail component seems to share the region with \HAEBE X-rays,
while soft X-rays from O stars do not.
Our results may suggest that the hard X-ray tail and \HAEBE X-ray activity have something 
in common (e.g. high energy activity of fossil magnetic fields) though samples are quite limited.

\section{Conclusion}

X-ray emission from \HAEBEs has long been a puzzle.
This is partly because the proper X-ray emission mechanism is not known for these
intermediate-mass young stars and partly because satisfying observing results
have not been obtained.
By combining our \ASCA analyses with previous \ROSAT result,
we find that the X-ray emission from \HAEBEs may originate in magnetic activity,
which begins in the earliest phase of stars and continues to the MS stage.
We list the results and implications below.

\begin{enumerate}
\item In the analyses of the \ASCA data, 11 X-ray counterparts are detected.
	The X-ray luminosity ranges between log \LX = 30 -- 32 \UNITLUMI,
	which is significantly larger than the typical luminosity of low-mass stars.
\item Four (or five if we include TYHD) sources of the 11 detected sources (36$-$45\%) exhibited time variation.
	Three \HAEBEs showed prominent variations with longer decay time scale (10$-$60 ksec).
\item X-ray spectra are reproduced by absorbed one or two temperature
	thin-thermal (MeKaL) models.
	The plasma temperature ranges between \KT $\sim$1$-$5 keV, which is significantly
	higher than those of MS OB stars (\KT $<$1 keV).
\item \HAEBEs have similar characteristics to low-mass YSOs in plasma temperature and time variability,
but not in the \LX/\Lbol ratio and $v_{rot}$ sin $i$ dependence.
        X-ray emission might originate from magnetic activity, but the activity might not be
        produced by solar-type magnetic dynamo.
\item If \NH is an indicator of stellar age, 
	younger \HAEBEs tend to have hotter plasma.
        While \LX goes down below 10$^{30}$ \UNITLUMI after the stellar age of
	$\sim$10$^{6}$-year-old.
	We propose that X-ray activity of \HAEBEs might be driven by magnetic interaction
	of a fossil field between the star and its accretion disk.
\end{enumerate}

For statistical analyses of young intermediate and high-mass stars, \CHANDRA will provide sophisticated data with little source contamination.
Time profiles with high photon statistics can be obtained by \XMM.
While the micro-calorimeter onboard Astro-E2 will provide a powerful tool for 
performing high resolution spectroscopy in the hard X-rays above 2 keV.

\acknowledgements
We greatly appreciate useful comments to Thierry Montmerle, Beate Stelzer,
Nicholas E. White and Michael F. Corcoran.
We also thank the referee for valuable comments that have led to substantial improvements in the
analysis and presentation of the paper.
This work is performed while the author (K.H.) held awards by the Japan Society for the Promotion of Science for Young 
Scientists (JSPS),  National Space Development Agency of Japan (NASDA) and 
National Research Council Research Associateship Award at NASA/GSFC.
The data in the article is obtained from the High Energy Astrophysics Division Online Service, 
provided by Institute of Space and Astronautical Science (ISAS) and the High Energy 
Astrophysics Science Archive Research Center Online Service, 
provided by the NASA/Goddard Space Flight Center.

\bibliographystyle{apj}
\bibliography{inst,sci,scibook}

\clearpage

\begin{deluxetable}{lrrrrrrclccrcc}
\tablecolumns{14}
\tablewidth{0pc}
\tabletypesize{\scriptsize}
\tablecaption{Stellar Parameters\label{tbl:starpar}}
\tablehead{
\colhead{Object}&\multicolumn{3}{c}{R.A. (2000)}&\multicolumn{3}{c}{Dec. (2000)}&\colhead{$d$}&\colhead{Sp. T}&\colhead{log \Teff}&\colhead{log \Lbol}&\colhead{\AV}&\colhead{$v_{wind}$}&\colhead{$v_{rot}$ sin $i$}\\
&(h&m&s)&(\DEGREE&\ARCMIN&\ARCSEC)&\colhead{(pc)}&\colhead{}&\colhead{(\DEGREEKELV)}
&\colhead{(\UNITSOLARLUMI)}&\colhead{}&\colhead{(\UNITVEL)}&\colhead{(\UNITVEL)}}
\startdata
BD +30 549&3&29&19.8&31&24&57&350&B8&4.08&1.1&1.9&\nodata&\nodata\\
V892 Tau&4&18&40.6&28&19&17&160&A6&\nodata&\nodata&4.1&250&\nodata\\
AB Aur&4&55&45.8&30&33&4&144&A0&3.98&1.7&0.5&260&140\\
V372 Ori&5&34&47.0&$-$5&34&15&460\tablenotemark{a}&B9.5+A0.5&3.93&2.2&0.5&\nodata&125\\
HD 36939&5&34&55.3&$-$5&30&22&460\tablenotemark{a}&B8--9&4.05&1.9&0.5&\nodata&275\\
HD 245185&5&35&9.6&10&1&52&400&A2&3.96&1.3&0.1&\nodata&150\\
LP Ori&5&35&9.8&$-$5&27&53&460\tablenotemark{a}&B2&4.29&3.1&0.7&\nodata&100\\
MR Ori&5&35&17.0&$-$5&21&46&460\tablenotemark{a}&A2&3.93&1.9&1.6&\nodata&\nodata\\
V361 Ori&5&35&31.4&$-$5&25&16&460\tablenotemark{a}&B1.5 or B4&4.14&2.6&0.4&\nodata&50\\
T Ori&5&35&50.0&$-$5&28&42&460&A3&3.93&1.6&1.1&\nodata&100\\
V380 Ori&5&36&25.4&$-$6&42&58&460&B9--A0&3.97&1.9&1.4&260&200\\
BF Ori&5&37&13.3&$-$6&35&1&430&A5--6&3.90&0.1&0.3&\nodata&100\\
MWC 120&5&41&2.3&$-$2&43&1&500&A2&3.95&1.3&0.03&\nodata&120\\
VY Mon&6&31&7.0&10&26&5&800&O9$-$B8&4.05&2.9&7.4&\nodata&\nodata\\
VY Mon G2\tablenotemark{b}&6&31&8.2&10&26&1&800&A0&(3.99)&4.0&1.6&\nodata&\nodata\\
V590 Mon&6&40&41.3&9&48&1&800&B7&4.09&2.6&0.6&\nodata&\nodata\\
LkH$\alpha$ 218&7&2&42.3&$-$11&26&10&1150&B9&4.03&2.1&1.5&\nodata&\nodata\\
Z CMa&7&3&43.2&$-$11&33&6&1150&F6&3.80&3.4&2.8&\nodata&$<$130\\
LkH$\alpha$ 220&7&4&5.4&$-$11&26&0&1150&B5&4.19&2.4&\nodata&\nodata&\nodata\\
HD 76534&8&55&8.7&$-$43&28&0&830&B2&4.34&3.9&1.2&\nodata&110\\
HD 97048&11&8&3.3&$-$77&39&17&180&B9--A0&4.00&1.6&1.2&\nodata&140\\
HD 97300&11&9&50.0&$-$76&36&48&188&B9&(4.03)&1.5&1.3&\nodata&\nodata\\
HD 104237&12&0&5.1&$-$78&11&35&116&A4&3.93&1.6&0.3&500&\nodata\\
IRAS 12496$-$7650&12&53&16.1&$-$77&7&2&200&A--F&(3.91)&1.7&12&\nodata&\nodata\\
HR 5999&16&8&34.3&$-$39&6&18&210&A5--7&3.90&1.9&0.5&100&180\\
HD 147889&16&25&24.3&$-$24&27&57&140&B2&4.34&3.3&3.3&\nodata&\nodata\\
Hen 3$-$1191&16&27&14.2&$-$48&39&28&\nodata&B0&\nodata&\nodata&\nodata&\nodata&\nodata\\
V921 Sco&16&59&6.9&$-$42&42&8&500&B4--5&(4.14)&3.0&7.1&500&\nodata\\
MWC 297&18&27&39.6&$-$3&49&52&250&B1.5&(4.38)&4.5&8&380&350\\
HD 176386&19&1&38.9&$-$36&53&27&122&B9&4.03&1.7&0.6&\nodata&\nodata\\
TY CrA&19&1&40.8&$-$36&52&34&130&B7--9&(4.07)&1.8&1.0&\nodata&10\\
R CrA&19&1&53.7&$-$36&57&8&130&B8&(4.06)&2.1&1.9&\nodata&\nodata\\
T CrA&19&1&59.0&$-$36&58&0&130&F0&(3.86)&0.9&1.7&\nodata&40\\
HD 200775&21&1&36.9&68&9&48&429&B2.5&4.31&3.9&1.9&280&60\\
MWC 1080&23&17&26.1&60&50&43&2200&A0--3&(3.96)&3.9&4.4&400&200\\ \hline 
%%\multicolumn{13}{l}{Sources with the spectral type less than G}\\ \hline
%%CoKu Tau 1&4&18&51.5&28&20&28&160&2&M2&(3.55)&\nodata&2.1\\
%%IX Ori&5&34&40.8&$-$5&22&43&460\tablenotemark{a}&2&K&\nodata&\nodata&0.5\\
%%YZ Ori&5&34&54.0&$-$5&3&30&460\tablenotemark{a}&2&K5&3.64&0.0&0.2\\
%%T Cha &11&57&13.5&$-$79&21&32&66&2&G2&3.77&0.1&1.6\\
EC 95&18&29&57.9&1&12&47&310&K0--K4&(3.65)&1.8&36&\nodata&\nodata\\
%%S CrA&19&1&8.5&$-$36&57&20&130&2&K6&(3.62)&{$-$}0.1&2.8\\
\enddata
\tablecomments{
Sp.T: spectral type, \Teff: parentheses show average temperature of their spectral type.
References$-$ \citet{Ancker1997,Ancker1998}, \citet{The1994}, \citet{Preibisch1998}, \citet{Nisini1995}, \citet{Benedettini1998}, \citet{Grady1993}, \citet{Boehm1995}, \citet{Drew1997}, \citet{Casey1998}, \citet{Ancker1998}, \citet{Corporon1999}.
}
\tablenotetext{a}{distance to the Ori OB1 association}
\tablenotetext{b}{We refer the relative position from VY Mon to \citet{Weintraub1990}.}
\end{deluxetable}

\begin{deluxetable}{lllrlrr}
\tablecolumns{7}
\tablewidth{0pc}
\tabletypesize{\scriptsize}
\tablecaption{\ASCA Observation Log\label{tbl:ascaobslog}}
\tablehead{
\colhead{Seq. ID}&\colhead{Abbrevation}&\colhead{Date}&\colhead{Ontime}&
\colhead{SIS Mode}&\multicolumn{2}{c}{Exposure}\\ 
\cline{6-7}\\
\colhead{}&\colhead{}&\colhead{}&\colhead{}&\colhead{}&\colhead{SIS}&\colhead{GIS}\\ 
\colhead{}&\colhead{}&\colhead{}&\colhead{(ksec)}&\colhead{}&\colhead{(ksec)}&\colhead{(ksec)}}
\startdata
23021000&SVS13&1995 Aug 30&211.0&\nodata&\nodata&93.0\\
23009000&IC359&1995 Sep 2&104.6&4F/B LD 0.55&27.4&37.8\\
23042000&SU Aur&1995 Feb 25&114.5&2F/B&40.8&45.0\\
20004000&Ori Trap1&1993 Aug 30&51.4&4F/B&19.3&18.0\\		
%% 25032000&Ori Trap2&1997 Sep 23&97.9&1F/B&45.3&8.2\\
%% 25032010&Ori Trap3&1997 Sep 30&110.3&1F/B&7.9&49.8\\
%% 25032020&Ori Trap4&1998 Sep 20&113.7&1F/B&40.4&35.9\\
%% 25035000&Iota Ori1&1997 Sep 21&149.0&\nodata&\nodata&72.5\\
%% 25036000&Iota Ori2&1997 Oct 6&73.0&\nodata&\nodata&24.6\\
21005000&L1641N&1994 Mar 13&97.5&\nodata&\nodata&34.6\\
21025000&Lamb Ori&1994 Mar 12&51.3&2/1F&21.8&22.4\\
23019000&Ori OB1&1995 Oct 4&79.8&\nodata&\nodata&28.6\\
26007000&PSRJ0631&1998 Oct 16&157.2&\nodata&\nodata&81.6\\
25015000&15 Mon&1997 Oct 18&96.9&1F LD 0.48&36.6&36.1\\
23014000&Z CMa&1995 Mar 31&79.6&1F&30.5&37.6\\
55030000&Vela Shrap&1997 May 9&39.6&\nodata&\nodata&12.3\\
21009000&ChamI I&1994 May 14&39.3&4F/B&8.5&15.2\\
%% 21009010&ChamI II&1994 May 14&57.8&\nodata&\nodata&13.9\\
%% 21009020&ChamI III&1994 May 15&46.7&\nodata&\nodata&16.8\\
%% 21009030&ChamI IV&1994 May 15&44.7&4F/B&12.6&16.4\\
27017000&ChamI V&1999 Aug 7&258.9&2F&82.4&75.0\\
%% 27002000&DC300-17&1999 Mar 8&69.1&1F&62.5&66.2\\
23003000&HD104237&1995 Apr 3&67.6&1F&29.9&27.1\\
24000000&ChamII&1996 Mar 10&69.1&\nodata&\nodata&24.9\\
24008000&Lupus3&1996 Feb 22&97.6&4F/B LD 0.7&18.2&41.3\\
20015010&Rho-Oph&1993 Aug 20&73.6&\nodata&\nodata&41.1\\
55003060&Gal R 1&1997 Sep 3&31.9&\nodata&\nodata&9.7\\
54004020&Gal R 2&1996 Aug 31&27.6&4F/B&12.2&	11.8\\
21007000&MWC297 1&1994 Apr 8&49.9&1F&10.5&10.3\\
21007010&MWC297 2&1994 Apr 12&24.9&1F&5.3&5.0\\
21007020&MWC297 3&1994 Apr 12&45.4&1F&12.9&12.3\\
21001000&R CrA 1&1994 Apr 4&90.3&4F/B&38.4&39.4\\
21002000&R CrA 2&1994 Apr 8&97.5&4F/B&35.5&36.9\\
24016000&R CrA 3&1996 Apr 5&160.2&1F&65.2&48.8\\
24016010&R CrA 4&1996 Oct 18&30.8&1F&10.6&11.7\\
66017000&R CrA 5&1998 Apr 19&57.6&1F&17.1&16.2\\
26018000&R CrA 6&1998 Oct 19&108.3&1F&31.5&34.1\\
23029000&HD200775&1995 Nov 17&110.0&1F LD 0.55&46.0&48.7\\
21006000&MWC1080&1993 Dec 8&86.0&1F&36.1&38.4\\ \hline
25021000&Serp&1997 Apr 13&246.2&4F LD 0.7&87.4&89.9\\
\enddata
\tablecomments{
Ontime: duration of observations,
SIS Mode: number of CCD chips working (4/2/1), Data Format (Faint/Bright), Level Discri (Unit keV), Left values on slashes are CCD modes during the high telemetry mode and right values are modes during the medium and low telemetry modes.}
\end{deluxetable}

\begin{deluxetable}{lllrlc}
\tablecolumns{6}
\tablewidth{0pc}
\tabletypesize{\scriptsize}
\tablecaption{\ROSAT Data Log\label{tbl:rosatobslog}}
\tablehead{
\colhead{Object}&\colhead{Obs ID}&\colhead{Date}&\colhead{Exp.}&\colhead{Detector}&\colhead{Usage}\\ 
\colhead{}&\colhead{}&\colhead{}&\colhead{(ksec)}&\colhead{}&\colhead{}}
\startdata
HD 245185&RH202047N00&1995 Mar 16&9.4&HRI&U\\
%%V380 Ori&RH900010A00&1991 Mar 22&11.0&HRI\\
BF Ori	&RH900010A01&1992 Mar 14&4.8&HRI&U\\
MWC 120&RP900189N00&1991 Sep 19&24.2&PSPC&U\\
VY Mon  &RP900355N00&1992 Oct 6&5.6&PSPC&ID\\
	&RP900355A01&1993 Mar 14&4.6&PSPC&ID\\
	&RP900355A02&1993 Sep 16&5.4&PSPC&ID\\
	&RP900355A03&1994 Mar 29&5.1&PSPC&ID\\
V590 Mon&RH200130A00&1991 Mar 20&19.6&HRI&U\\
LkH$\alpha$ 218&RP201011N00&1993 Apr 22&19.7&PSPC&U\\
LkH$\alpha$ 220&RH202153N00&1996 Apr 6&33.5&HRI&U\\ 
MWC 297 &RH202048N00&1995 Sep 28&1.1&HRI&ID\\
HD 200775&RH202319N00&1997 Aug 11&4.8&HRI&ID\\
\enddata
\tablecomments{U: used for upper limit measurement (See also 
Table \ref{tbl:imgnondet}). ID: used for source identification.}
\end{deluxetable}

\begin{deluxetable}{lcll}
\tablecolumns{4}
\tablewidth{0pc}
\tabletypesize{\scriptsize}
\tablecaption{Selected Regions\label{tbl:reg}}
\tablehead{
\colhead{Object}&\colhead{Det.}&\multicolumn{2}{c}{Region}\\
\cline{3-4}\\
\colhead{}&\colhead{}&\colhead{Source}&\colhead{Background}}
\startdata
V892 Tau&S	&c(3)\tablenotemark{a}&sym V1023 Tau\\ 
V380 Ori&G	&c(3)	&a(3/6)	\\
VY Mon  &G	&e(4.5/2.5)&$sf$ (sym optical axis)\\
HD 104237&S	&c(2.5)	&whole chip excluding $sr$\\
	&G	&c(3)	&a(3/6)	\\
IRAS 12496$-$7650&G	&c(3)	&a(3/6)	\\
HR 5999	&S	&e(0.66/0.25)&sym HR 6000\\
V921 Sco&G	&c(3)	&a(3/6)	\\
MWC 297	&S	&c(3)	&a(3/4.4)	\\
	&G	&c(3)	&a(3/6)	\\
TY CrA	&S	&e(0.74/0.53)&sym HD 176386\\
HD 176386&S	&e(0.64/0.42)&sym TY CrA\\
~TYHD\tablenotemark{b}&G&c(2.5)&sym the R CrA protostar cluster\\
HD 200775&S&c(2.5)&$sf$\\
	&G&c(3)&sym the SE source\\
MWC 1080&S&c(2.5)&$sf$\\
	&G&c(3)&a(3/6)\\ \hline
%% T Cha   &S	&c(3)	&$sf$	\\
%%	&G	&c(3)	&a(3/6)	\\
%% S CrA	&S	&c(1.5) &sym CrA~1\\
EC 95	&S	&c(3)	&b(9/3)	\\
	&G	&c(3)	&a(3/6)	\\
\enddata
\tablecomments{
c($num$): $num$ arcmin radius circle centered on a target,
e($num1$/$num2$): ellipse with $num1$ arcmin long and $num2$ arcmin short axes,
a($num1$/$num2$): annular circle with $num1$ arcmin inner and $num2$ arcmin outer radii,
b($num1$/$num2$): $num1$ arcmin $\times$ $num2$ arcmin box,
sym $sth$: symmetrical region with respect to $sth$,
$sr$: source region, $sf$: source free region}
\tablenotetext{a}{Including V410 X-ray 7}
\tablenotetext{b}{GIS data merged with TY CrA and HD 176386}
\end{deluxetable}

\begin{deluxetable}{llrrrrrrccrlrlr}
\tablecolumns{15}
\tablewidth{0pc}
\tabletypesize{\scriptsize}
\tablecaption{\ASCA Detected Sources\label{tbl:imgdet}}
\tablehead{
\colhead{Object}&\colhead{Obs. ID}&\multicolumn{7}{c}{Detected Position}&\colhead{}&\multicolumn{5}{c}{Net Count}\\
\cline{3-9}\cline{11-15}\\
\colhead{}&\colhead{}&\multicolumn{3}{c}{R.A. (2000)}&\multicolumn{3}{c}{Dec. (2000)}&\colhead{Det.}&\colhead{}&\multicolumn{2}{c}{SIS}&\multicolumn{2}{c}{GIS}&\colhead{Total}\\
\colhead{}&\colhead{}&\colhead{(h}&\colhead{m}&\colhead{s)}&\colhead{(\DEGREE}&\colhead{\ARCMIN}&\colhead{\ARCSEC)}&\colhead{}&\colhead{}&\colhead{(cnts)}
&\colhead{(\%)}&\colhead{(cnts)}&\colhead{(\%)}&\colhead{(cnts)}}
\startdata
V892 Tau&IC359&4&18&39.8&28&19&16&S&&722&(59)&\nodata&(\nodata)&722\\
V380 Ori&L1641N&5&36&23.8&$-$6&43&18&G2&&\nodata&(\nodata)&109\tablenotemark{b}&(33)&109\\
VY Mon (/G2)&PSRJ0631&6&31&3.1&10&26&46&G&&\nodata&(\nodata)&338&(36)&338\\
HD 104237&HD104237&12&0&6.2&$-$78&11&29&S&&2427&(84)&916&(74)&3343\\
IRAS 12496$-$7650&ChamII&12&53&16.5&$-$77&7&6&G&&\nodata&(\nodata)&100&(31)&100\\
HR 5999&Lupus3&16&8&34.9&$-$39&6&12&S&&143&(48)&\nodata&(\nodata)&143\\
V921 Sco&Gal R 2&16&59&8.7&$-$42&42&30&G&&\nodata&(\nodata)&151&(38)&151\\
MWC 297&MWC297 1&18&27&37.9&$-$3&49&36&S&&353&(62)&257&(67)&610\\
      &MWC297 2&\multicolumn{7}{c}{}&&1023&(79)&904&(87)&1927\\
      &MWC297 3&\multicolumn{7}{c}{}&&528&(64)&349&(70)&877\\
TY CrA	&R CrA 1&19&1&40.8&$-$36&52& 31&S&&524&(67)&\nodata&(\nodata)&524\\
	&R CrA 4&\multicolumn{7}{c}{}&&75\tablenotemark{a}&(56)&\nodata&(\nodata)&75\\
	&R CrA 6&&&&&&&&&437&(71)&\nodata&(\nodata)&437\\
HD 176386&R CrA 1&19&1& 38.7&$-$36&53&18&S&&227&(49)&\nodata&(\nodata)&227\\
	&R CrA 4&\multicolumn{7}{c}{}&&45\tablenotemark{a}&(52)&\nodata&(\nodata)&45\\
	&R CrA 6&&&&&&&&&228&(49)&\nodata&(\nodata)&228\\
~TYHD&R CrA 1&\multicolumn{7}{c}{}&&\nodata&(\nodata)&1488&(78)&1488\\
	&R CrA 2&\multicolumn{7}{c}{}&&\nodata&(\nodata)&676&(64)&676\\
	&R CrA 3&\multicolumn{7}{c}{}&&\nodata&(\nodata)&1585\tablenotemark{c}&(78)&1585\\
	&R CrA 4&\multicolumn{7}{c}{}&&\nodata&(\nodata)&510&(73)&510\\
	&R CrA 5&\multicolumn{7}{c}{}&&\nodata&(\nodata)&543&(91)&543\\
	&R CrA 6&\multicolumn{7}{c}{}&&\nodata&(\nodata)&1029&(68)&1029\\
HD 200775&HD200775&21&1&34.9&68&10&1&S&&3361&(85)&1641&(69)&5002\\
MWC 1080&MWC1080&23&17&24.0&60&50&43&S&&273&(33)&185&(27)&458\\ \hline
EC 95&Serp&18&29&58.3&1&12&43&S&&3216&(75)&3679&(70)&6895\\
\enddata
\tablecomments{
Det.: detectors measuring source positions (S: SIS, G: GIS),
Net Count: background subtracted source counts, Parentheses: 
source event ratio defined as ``source counts / (source + background) counts'',
Total: net counts of SIS + GIS, 
TYHD: GIS data merged with TY CrA and HD 176386.}
\tablenotetext{a}{s0 data}
\tablenotetext{b}{g2 data}
\tablenotetext{c}{g3 data}
\end{deluxetable}

\begin{deluxetable}{llllcrlrcrrrrr}
\tablecolumns{14}
\tablewidth{0pc}
\rotate
\tabletypesize{\scriptsize}
\tablecaption{\ASCA Non-detected Sources\label{tbl:imgnondet}}
\tablehead{
\colhead{Object}&\multicolumn{6}{c}{\ASCA}&&\multicolumn{5}{c}{\ROSAT}\\
\cline{2-7}\cline{9-13}\\
\colhead{}&\colhead{Obs. ID}&\multicolumn{2}{c}{Flux upper limit}&\colhead{Det.}&\colhead{Log \LX}&\colhead{Note}&\colhead{}&\colhead{Ref.}&\multicolumn{2}{c}{Flux}&\colhead{Log \LX}&\colhead{Note}&\colhead{Log $\frac{L_X}{L_{bol}}$}\\
\colhead{}&\colhead{}&\colhead{Photon}&\colhead{Energy}&\colhead{}&\colhead{}&\colhead{}&\colhead{}&\colhead{}&\colhead{Photon}&\colhead{Energy}&\colhead{}&\colhead{}}
\startdata
BD +30 549&SVS13&2.1~10$^{-2}$&1.1~10$^{-12}$&G&$<$31.2&con.&&1&\nodata&\nodata&$<$29.8&\nodata&$<-$4.9\\
AB Aur&SU Aur&3.1~10$^{-2}$&7.4~10$^{-13}$&S&$<$30.3&con.&&2&\nodata&\nodata&29.5&\nodata&$-$5.7\\
V372 Ori&Ori Trap1&3.5~10$^{-2}$&1.3~10$^{-12}$&G&$<$31.5&\nodata&&3&\nodata&\nodata&30.3&\nodata&$-$5.5\\
HD 36939&Ori Trap1&4.6~10$^{-2}$&1.7~10$^{-12}$&G&$<$31.6&\nodata&&3&\nodata&\nodata&$<$29.9&\nodata&$<-$5.6\\
HD 245185&Lamb Ori&1.6~10$^{-2}$&3.3~10$^{-13}$&S&$<$30.8&\nodata&&4&$<$2.4~10$^{-3}$&$<$9.7~10$^{-14}$&$<$30.3&\nodata&$<-$4.6\\
LP Ori&Ori Trap1&3.6~10$^{-1}$&9.2~10$^{-12}$&S&$<$32.4&con.&&3&\nodata&\nodata&30.1&\nodata&$-$6.6\\
MR Ori&Ori Trap1&2.7&9.4~10$^{-11}$&S&$<$33.4&con.&&3&\nodata&\nodata&$<$29.9&\nodata&$<-$5.2\\
V361 Ori&Ori Trap1&9.8~10$^{-1}$&2.3~10$^{-11}$&S&$<$32.8&con.&&3&\nodata&\nodata&30.9&\nodata&$-$5.4\\
T Ori&Ori Trap1&6.5~10$^{-2}$&1.6~10$^{-12}$&S&$<$31.6&con.&&3&\nodata&\nodata&29.9&\nodata&$-$5.3\\
BF Ori&L1641N&\nodata&5.8~10$^{-13}$&G&$<$31.1&ano.&&4&$<$2.2~10$^{-3}$&$<$6.7~10$^{-14}$&$<$30.2&\nodata&$<-$3.5\\
MWC 120&Ori OB1&9.5~10$^{-3}$&3.1~10$^{-13}$&G&$<$31.0&\nodata&&4&$<$2.7~10$^{-3}$&$<$3.3~10$^{-14}$&$<$30.0&\nodata&$<-$4.9\\
V590 Mon&15 Mon&4.8~10$^{-2}$&1.2~10$^{-12}$&S&$<$32.0&\nodata&&4&4.2~10$^{-3}$&2.4~10$^{-13}$&31.3&\nodata&$-$4.8\\
LkH$\alpha$ 218&Z CMa&9.9~10$^{-2}$&4.4~10$^{-12}$&G&$<$32.8&\nodata&&4&$<$3.7~10$^{-2}$&$<$1.2~10$^{-12}$&$<$32.3&con.&$<-$3.4\\
Z CMa&Z CMa&1.9~10$^{-2}$&8.9~10$^{-13}$&S&$<$32.1&\nodata&&2&\nodata&\nodata&31.1&\nodata&$-$5.8\\
LkH$\alpha$ 220&Z CMa&8.5~10$^{-3}$&2.8~10$^{-13}${$^{ \dagger}$}&G&$<$31.6&con.&&4&$<$7.0~10$^{-4}$&$<$2.2~10$^{-14}$&$<$30.5&\nodata&$<-$5.4\\
HD 76534&Vela Shrap&1.0~10$^{-2}$&4.3~10$^{-13}$&G&$<$31.6&\nodata&&5&\nodata&\nodata&\nodata&\nodata&$<-$5.9\\
HD 97048&ChamI V&\nodata&1.8~10$^{-13}$&S&$<$29.8&ano.&&2&\nodata&\nodata&29.0&\nodata&$-$6.2\\
HD 97300&ChamI I&5.1~10$^{-2}$&2.3~10$^{-12}$&G&$<$31.0&con.&&6&\nodata&\nodata&29.0&\nodata&$-$6.1\\
HD 147889&Rho-Oph&\nodata&1.7~10$^{-12}$&G&$<$30.6&ano.&&7&\nodata&\nodata&$<$27.5&\nodata&$<-$9.4\\
Hen 3-1191&Gal R 1&2.3~10$^{-2}$&7.5~10$^{-13}${$^{ \dagger}$}&G&\nodata&\nodata&&5&\nodata&\nodata&\nodata&\nodata&\nodata\\
R CrA&R CrA 1&1.4~10$^{-1}$&5.2~10$^{-12}$&S&$<$31.0&con.&&2&\nodata&\nodata&$<$28.7&\nodata&$<-$7.0\\
T CrA&R CrA 1&4.0~10$^{-2}$&1.5~10$^{-12}$&S&$<$30.5&con.&&2&\nodata&\nodata&$<$28.5&\nodata&$<-$6.0\\
\enddata
\tablecomments{
{\it The column of} \ASCA, 
flux upper limit (0.5--10 keV), Photon [\UNITCPS]:
3$\sigma$ level or count rate of source contamination estimated from a 40\ARCSEC and 80\ARCSEC square regions for the SIS and GIS,
respectively.
Energy [\UNITFLUX]: unabsorbed energy flux upperlimit
calculated using the PIMMS package, assuming thin-thermal plasma with \KT~$\sim$2 keV and \NH converted from the \AV,
Log \LX [\UNITLUMI] (0.5--10 keV): absorption corrected X-ray luminosity,
Det.: detectors measuring count rates (S: SIS, G: GIS),
Sources marked as ``con.'' have source contamination,
Sources marked as ``ano.'' are identified as the other objects. $^{ \dagger}$ assuming no absorption (\NH = 0.0 \UNITNH).
{\it The column of} \ROSAT, 
ref.: reference of log \LX, 1: \citet{Preibisch1997}, 2: \citet{Zinnecker1994}, 3: \citet{Gagne1995}, 
4: this work, 5: no pointing observation, 6: \citet{Feigelson1993},
7: \citet{Casanova1995},
Flux (0.1--2.4 keV): Photon [\UNITCPS], Energy [\UNITFLUX], absorption corrected energy flux 
assuming thin-thermal plasma of \KT~$\sim$2 keV and \NH converted from \AV,
Log \LX [\UNITLUMI] (0.1--2.4 keV): absorption corrected X-ray luminosity.
Sources marked as ``con.'' have source contamination.}
\end{deluxetable}

\begin{deluxetable}{llcccrlc}
\tablecolumns{8}
\tablewidth{0pc}
\tabletypesize{\scriptsize}
\tablecaption{Result of \ASCA Timing Analysis\label{tbl:curvefit}}
\tablehead{
\colhead{Object}&\colhead{St.}&\colhead{Det.}&\colhead{Bin}&\multicolumn{4}{c}{Constant Fittings}\\
\cline{5-8}\\
\colhead{}&\colhead{}&\colhead{}&\colhead{}&\colhead{Mean}&
\multicolumn{2}{c}{Reduced $\chi^{2} (d.o.f)$}&\colhead{Var.}\\
\colhead{}&\colhead{}&\colhead{}&\colhead{(sec)}&\colhead{(\UNITCPS)}
&\colhead{}&\colhead{}&\colhead{}}
\startdata
V892 Tau&&S&2048&1.5~10$^{-2}$&1.42&(27)&n\\
V380 Ori&&G2&2048&2.4~10$^{-3}$&0.94&(33)&n\\
VY Mon (/G2)&&G&2048&8.9~10$^{-3}$&1.69&(70)&y\\
HD 104237&&SG&2048&2.8~10$^{-2}$&2.62&(31)&y\\
IRAS 12496$-$7650&&G&2048&1.6~10$^{-3}$&1.00&(30)&n\\
HR 5999&&S&2048&3.4~10$^{-3}$&1.08&(21)&n\\
V921 Sco&&G&2048&6.3~10$^{-3}$&0.51&(11)&n\\
MWC 297&1&SG&2048&1.5~10$^{-2}$&0.25&(10)&n\\
&2&SG&2048&9.0~10$^{-2}$&4.51&(7)&y\\
&3&SG&2048&1.6~10$^{-2}$&4.56&(13)&y\\
TY CrA&1&S&2048&6.4~10$^{-3}$&0.79&(34)&n\\
&4&S&2048&5.8~10$^{-3}$&1.09&(12)&n\\
&6&S&2048&6.7~10$^{-3}$&0.90&(39)&n\\
HD 176386&1&S&2048&2.8~10$^{-3}$&0.83&(33)&n\\
&4&S&2048&4.4~10$^{-3}$&0.40&(12)&n\\
&6&S&2048&2.1~10$^{-3}$&1.21&(39)&n\\
TYHD&1&G&2048&1.4~10$^{-2}$&1.67&(36)&y\\
&2&G&2048&1.3~10$^{-2}$&1.42&(37)&n\\
&3&G&4096&4.0~10$^{-2}$&3.14&(33)&y\\
&4&G&2048&1.5~10$^{-2}$&1.79&(14)&n\\
&5&G&2048&1.8~10$^{-2}$&0.84&(21)&n\\
&6&G&2048&6.7~10$^{-3}$&0.90&(39)&n\\
HD 200775&&S1G\tablenotemark{a}&2048&2.1~10$^{-2}$&1.71&(47)&y\\
MWC 1080&&SG&4096&3.0~10$^{-3}$&0.79&(19)&n\\ \hline
EC 95&&SG&2048&1.7~10$^{-2}$&2.28&(92)&y\\
\enddata
\tablecomments{
St.: number of observation ID., Det.: detector 
(S: SIS, G: GIS, SG: SIS + GIS),
Bin.: binning time scales of light curves,
Mean: average count rates derived from constant fittings,
Var.: sources showing time variation above 96\% confidence level,
TYHD: GIS data merged with TY CrA and HD 176386.}
\tablenotetext{a}{SIS1 + GIS}
\end{deluxetable}

\begin{deluxetable}{llccrlrlrlrlccl}
\tablecolumns{15}
\tablewidth{0pc}
\tabletypesize{\tiny}
\tablecaption{Result of \ASCA Spectral Analysis\label{tbl:specfit}}
\tablehead{
\colhead{Object}&\colhead{St.}&\colhead{Det.}&\colhead{Model}&\multicolumn{2}{c}{\KT}&\multicolumn{2}{c}{$E.M.$}&
\multicolumn{2}{c}{\NH}&\multicolumn{2}{c}{Reduced $\chi^{2} (d.o.f.)$}&\colhead{Flux}&\colhead{\LX}&\colhead{\LX/\Lbol}}
\startdata
V892 Tau&&S&1T&2.0&(1.6--2.5)&54.0&(53.9--54.1)&1.2&(0.9--1.6)&0.99&(51)&1.22&\nodata&\nodata\\
&&S&2T&2.0&(fix)&54.0&(fix)&1.8&(fix)&0.87&(54)&1.02&31.0&--4.4\\
&&&&0.5&(0.3--0.7)&54.3&(53.9--55.4)&1.7&(1.2--2.8)&\nodata&&0.20&31.3\tablenotemark{b}&\nodata\\
V380 Ori&&G2&1T&3.2&(0.7--11.0)&54.1&(53.9--54.3)\tablenotemark{a}&0.0&(0.0--1.5)&1.31&(14)&0.59&31.2&--4.4\\
&&G2&1T{$^{F}$}&2.4&(1.3--6.1)&54.2&(54.0--54.4)&0.3&(fix)&1.25&(15)&0.68&31.2&--4.3\\
VY Mon (/G2)&LS&G&1T&2.9&(2.0--4.6)&55.1&(55.0--55.3)&0.6&(0.2--1.0)&1.40&(42)&1.21&32.2&--5.4\tablenotemark{d}\\
&LS&G&1T$^{G}$\tablenotemark{c}&2.2&(1.6--3.3)&55.2&(55.0--55.4)&0.8&(0.4--1.2)&1.23&(40)&1.18&32.2&--5.4\tablenotemark{d}\\
&HS&G&1T&6.0&(3.8--12.3)&55.2&(55.1--55.3)&0.4&(0.1--0.8)&1.29&(45)&2.49&32.4&--4.1\tablenotemark{d}\\
&HS&G&1T{$^{LS}$}&20.3&(3.0--)&54.8&(54.7--55.2)&0.6&(0.0--2.5)&1.28&(45)&1.24&32.1&--4.4\tablenotemark{d}\\
HD 104237&&SG&1T&2.1&(\nodata)&53.3&(\nodata)&0.0&(\nodata)&3.82&(89)&\nodata&\nodata&\nodata\\
&&SG&2T&0.7&(0.6--0.7)&52.9&(52.8--53.1)&0.1&(0.0--0.2)&1.57&(87)&0.43&29.9&--5.2\\
&&&&4.0&(3.3--4.8)&53.1&(53.0--53.1)&0.1&(com.)&\nodata&&0.93&30.2&--4.9\\
IRAS 12496$-$7650&&G&1T&2.0&(0.9--7.2)&54.1&(53.4--55.3)&13&(6.2--24)&0.93&(14)&0.29&31.1&--4.2\\
HR 5999&&S&1T&0.9&(0.6--1.7)&53.9&(53.5--54.3)&0.7&(0.0--1.2)&1.15&(11)&0.37&30.9&--4.6\\
&&S&1T{$^{F}$}&1.6&(1.2--2.4)&53.5&(53.4--53.6)&0.1&(fix)&1.27&(12)&0.44&30.4&--5.1\\
V921 Sco&&G&1T&2.8&(1.1--9.1)&55.0&(54.6--56.1)&4.9&(2.3--14)&0.96&(18)&1.14&32.1&--4.5\\
MWC 297&1&SG&1T&3.4&(2.4--5.1)&54.2&(54.1--54.4)&2.4&(1.8--3.1)&0.92&(43)&1.18&31.3&--6.8\\
&2&SG&1T&6.5&(4.7--9.6)&54.9&(54.8--55.0)&2.5&(2.1--2.8)&1.04&(105)&9.23&32.1&--6.0\\
&3&SG&1T&3.4&(2.6--4.8)&54.2&(54.1--54.3)&1.9&(1.5--2.4)&0.89&(63)&1.31&31.3&--6.7\\
TY CrA&1&S&1T&1.8&(0.8--2.3)&53.4&(53.3--53.5)&0.3&(0.1--0.5)&1.02&(28)&0.76&30.4&--5.1\\
&4&S&1T{$^{F}$}&1.8&(1.2--3.4)&53.5&(53.3--53.7)&0.4&(fix)&1.33&(5)&0.89&30.5&--4.9\\
&6&S&1T&1.2&(1.0--1.4)&53.7&(53.5--53.8)&0.6&(0.3--0.9)&0.87&(25)&0.82&30.6&--4.8\\
HD 176386&1&S&1T&1.3&(1.0--1.8)&52.8&(52.7--52.9)&0.0&(0.0--0.2)&0.90&(16)&0.32&29.8&--5.5\\
&4&S&1T{$^{F}$}&1.0&(0.7--1.5)&53.0&(52.8--53.2)&0.0&(fix)&0.56&(4)&0.50&30.0&--5.3\\
&6&S&1T&1.9&(1.3--2.8)&52.9&(52.8--53.9)&0.0&(0.0--0.2)&1.93&(13)&0.39&29.8&--5.4\\
~TYHD&1&G&1T&2.1&(1.9--2.4)&53.6&(53.6--53.6)&0.0&(0.0--0.1)&1.87&(26)&1.86&\nodata&\nodata\\
&2&G&1T&1.7&(1.2--2.0)&53.6&(53.6--53.8)&0.0&(0.0--0.5)&0.66&(41)&1.74&\nodata&\nodata\\
&3LS&G3&1T&2.6&(2.1--3.1)&53.7&(53.6--53.7)&0.0&(0.0--0.1)&0.79&(65)&2.43&\nodata&\nodata\\
&3HS&G3&1T&3.0&(2.4--3.8)&53.8&(53.8--53.9)&0.0&(0.0--0.2)&0.88&(22)&3.87&\nodata&\nodata\\
&4&G&1T&2.5&(2.0--3.3)&53.6&(53.5--53.6)&0.0&(0.0--0.2)&1.18&(30)&1.92&\nodata&\nodata\\
&5&G&1T&2.1&(1.3--2.6)&53.7&(53.6--53.9)&0.0&(0.0--0.4)&0.98&(25)&2.26&\nodata&\nodata\\
&6&G&1T&1.7&(1.4--1.9)&53.5&(53.5--53.6)&0.0&(0.0--0.2)&1.03&(64)&1.49&\nodata&\nodata\\
HD 200775&&SG&1T&1.7&(1.6--1.9)&54.5&(54.5--54.6)&0.1&(0.0--0.2)&1.39&(211)&1.11&31.5&--6.0\\
&&SG&2T&0.5&(0.4--0.7)&55.2&(55.1--55.4)&0.9&(0.8--1.0)&1.15&(209)&0.59&32.1&--5.3\\
&&&&2.8&(1.9--3.8)&54.2&(54.1--54.5)&0.9&(com.)&\nodata&&0.47&31.3&--6.2\\
MWC 1080&&SG&1T&10.0&(3.8--)&55.1&(55.0--55.3)&1.0&(0.3--2.2)&1.02&(56)&0.26&32.3&--5.1\\
	&&SG&1T$^{Ge}$&7.2&(3.0--)&55.1&(54.9--55.3)&1.0&(0.3--2.2)&0.97&(54)&0.27&32.3&--5.1\\ \hline
EC 95&&SG&1T&3.9&(3.5--4.4)&54.5&(54.5--54.6)\tablenotemark{f}&2.3&(2.0--2.5)&1.20&(205)&1.97\tablenotemark{f}&31.7&--3.7\\
\enddata
\tablecomments{
St.: number of observation ID and/or status (HS: high or flare state, LS: low or quiescent state),
Det.: S: SIS, G: GIS, SG: SIS+GIS,
Model: 1T: 1-temperature, 2T: 2-temperature, $^{G/LS}$: adding the components of Gaussian/low state, $^{F}$: fixing \NH value,
\KT[keV], \EM[log, cm$^{-3}$], \NH[10$^{22}$ \UNITNH],
com.: common \NH values with the above column,
fix: parameter is fixed,
Parentheses show 90\% confidence level,
Reduced $\chi^{2}$: parentheses denote number of degree of freedom ($d.o.f$),
Flux [10$^{-12}$ \UNITFLUX]: observed flux (0.5--10 keV),
\LX[log, \UNITLUMI]: absorption corrected X-ray luminosity between 0.5--10 keV,
\LX/\Lbol[log],
TYHD: GIS data merged with TY CrA and HD 176386}.
\tablenotetext{a}{\NH is fixed in the error estimate}
\tablenotetext{b}{the distance to V410 X-ray 7}
\tablenotetext{c}{best-fit parameters of the Gaussian component 
[center energy, 5.1 (4.8$-$5.3) keV assuming line width of 0.0 eV.]}
\tablenotetext{d}{\Lbol of VY Mon G2}
\tablenotetext{e}{best-fit parameters of the Gaussian component 
[center energy, 6.4 (6.2$-$6.6) keV assuming line width of 0.0 eV.]}
\tablenotetext{f}{GIS value}
\end{deluxetable}

\begin{deluxetable}{llcllll}
\tablecolumns{7}
\tablewidth{0pc}
\tabletypesize{\scriptsize}
\tablecaption{Fitting Results of the Flare Light Curves\label{tbl:flarecurvefit}}
\tablehead{
\colhead{Object}&\colhead{Obs. ID}&\colhead{Det.}&\colhead{{\it I}$_{qui}$}
&\colhead{{\it I}$_{peak}$}&\colhead{$\tau_{ris}$}&\colhead{$\tau_{dec}$}\\
\colhead{}&\colhead{}&\colhead{}&\colhead{(\UNITCPS)}&\colhead{(\UNITCPS)}&\colhead{(10$^{4}$ sec)}&\colhead{(10$^{4}$ sec)}}
\startdata
MWC 297	&MWC297 2,3 &SG &2.3 (2.1--2.4) 10$^{-2}$&$>$ 1.2 10$^{-1}$ \tablenotemark{a}&\nodata &~~~5.6 (4.5--7.0) \\
VY Mon	&PSRJ0631&G&2.5 (1.7--2.9) 10$^{-3}$&$>$ 3.8 10$^{-3}$ \tablenotemark{a}&\nodata&~~~3.6 (2.0--7.4) \\
TYHD\tablenotemark{b}&R CrA3&G&3.6 (3.4--3.8) 10$^{-2}$&$\sim$ 7.5 10$^{-2}$&$\sim$ 1&~~~2.4 (1.3--8.0) \\
%%SSV 63E+W\tablenotemark{c}&\nodata&\nodata       &2.0 (1.6--2.6) 10$^{-2}$&~~~1.8 (1.6--1.9) 10$^{-2}$&~~~1.7&~~~1.2 (0.9--1.5) \\
\enddata
\tablecomments{
Det.: detector, S: SIS, G: GIS., 
$\tau_{dec}$: $e$-folding time fit by a constant and exponential decay model.
$\tau_{ris}$: the period between the constant level to the flux peak.}
\tablenotetext{a}{count rate at the observation start.}
\tablenotetext{b}{GIS data merged with TY CrA and HD 176386.}
%%\tablenotetext{c}{reference to Ozawa et al. (1999).}
\end{deluxetable}

\begin{deluxetable}{cp{4cm}|p{4cm}}
\tablecolumns{3}
\tablewidth{0pc}
\tabletypesize{\scriptsize}
\tablecaption{Outflow Activity\label{tbl:outflow}}
\tablehead{
\colhead{log \LX}&\multicolumn{2}{c}{Outflow Activity}\\ \cline {2-3}
\colhead{(\UNITLUMI)}&\colhead{Yes}&\colhead{No activity found}}
\startdata
Above 30 & V380 Ori$^{1,2}$, Z CMa$^2$, IRAS 12496-7650$^{3,5}$, HD 200775$^{2,4}$, MWC 1080$^{2,6}$, EC 95$^{7,}$\tablenotemark{a} & V892 Tau$^2$, HD 104237$^8$, HR 5999$^2$, MWC297$^3$\\ \hline
Below 30 & R CrA$^{2,3}$, T CrA$^2$ & T Ori$^2$, HD 76534$^2$, HD 97048$^3$\\
\enddata
\tablecomments{References: $^1$\citet{Corcoran1995}, $^2$\citet{Maheswar2002} (Table 2, column 8),
$^3$\citet{Lorenzetti1999} (Table 2, column 5), $^4$\citet{Fuente1998}, $^5$\citet{Hughes1989},
$^6$\citet{Poetzel1992}, $^7$\citet{Eiroa1989}, $^8$\citet{Knee1996}.
Sources with blanks in the columns of \citet{Maheswar2002} and \citet{Lorenzetti1999} are listed
in "no activity found".}
\tablenotetext{a}{proto-\HAEBE}
\end{deluxetable}

\clearpage

\clearpage

\clearpage

\clearpage

\clearpage

\clearpage

\clearpage

\clearpage

\clearpage

\clearpage

\clearpage

\clearpage

\clearpage

\clearpage

\clearpage

\clearpage

\clearpage

\clearpage

\end{document}